\pdfoutput=1 
\documentclass[fleqn,usenatbib]{mnras}

\usepackage{txfonts}

\usepackage[T1]{fontenc}
\usepackage{ae,aecompl}

\usepackage{graphicx}	

\usepackage{amsmath}	
\usepackage{amsfonts}   
\usepackage{amssymb}	
\usepackage{gensymb}    
\usepackage{pdflscape}  

\usepackage{changepage} 
\usepackage{enumitem} 
\usepackage{threeparttable} 

\def \aj {AJ}
\def \mnras {MNRAS}
\def \pasp {PASP}
\def \apj {ApJ}
\def \apjs {ApJS}

\def \aap {A\&A}

\def \araa {ARAA}

\newcommand{\Msolar} {$\mathrm{M_{\odot}}$}

\newcommand{\hii} {H{\scriptsize{II}} }
\newcommand{\kms} {$\mathrm{ km \; s^{-1}}\,$}

\def\lesssim{\mathrel{\hbox{\rlap{\hbox{\lower4pt\hbox{$\sim$}}}\hbox{$<$}}}}
\def\gtrsim{\mathrel{\hbox{\rlap{\hbox{\lower4pt\hbox{$\sim$}}}\hbox{$>$}}}}
\newcommand{\ang} {\r{A}$\,$}
\long\def\symbolfootnote[#1]#2{\begingroup%
\def\thefootnote{\fnsymbol{footnote}}\footnote[#1]{#2}\endgroup}

\title[Galactic Wolf-Rayet stars with \textit{Gaia} DR2 I]{Unlocking Galactic Wolf-Rayet stars with \textit{Gaia} DR2 I: Distances and absolute magnitudes}

\author[Rate \& Crowther]{
Gemma Rate,\thanks{garate1@sheffield.ac.uk} Paul A. Crowther
\\
Department of Physics and Astronomy, University of Sheffield, Sheffield, S3 7RH, UK\\
}

\date{Accepted XXX. Received YYY; in original form ZZZ}

\pubyear{2019}
\begin{document}
\label{firstpage}
\pagerange{\pageref{firstpage}--\pageref{lastpage}}
\maketitle

\begin{abstract}
  We obtain distances to 383 Galactic Wolf-Rayet (WR) stars from \textit{Gaia} DR2 parallaxes and Bayesian methods, with a prior based on \hii regions and dust extinction. Distances agree with those from Bailer-Jones et al. for stars up to 2 kpc from the Sun, though deviate thereafter due to differing priors,
   leading to modest reductions in luminosities for recent WR spectroscopic results. We calculate visual and  K-band absolute magnitudes, accounting for dust extinction contributions and binarity, and identify 187 stars with reliable  absolute magnitudes. For WR and O stars within 2 kpc, we find a WR/O ratio of 0.09. The distances are used to generate absolute magnitude calibrations and obtain the \textit{Gaia} colour magnitude diagram for WR stars.
   Average $v^{\rm WR}$-band absolute magnitudes for WN stars range from --3.6 mag (WN3--4) to --7.0 mag (WN8--9ha), and --3.1 (WO2--4) to --4.6 mag (WC9), with standard deviations of $\sim$0.6 mag. Using \hii region scale heights, we identify 31 WR stars at large (3$\sigma$, |z|$\geq$156 pc) distances from the mid-plane as potential runaways  accounting for the Galactic warp, of which only 4 involve WN8--9 stars, contrary to previous claims.
\end{abstract}





\section{Introduction}
			

\begin{figure*}
	\centering
	\includegraphics[width=\linewidth]{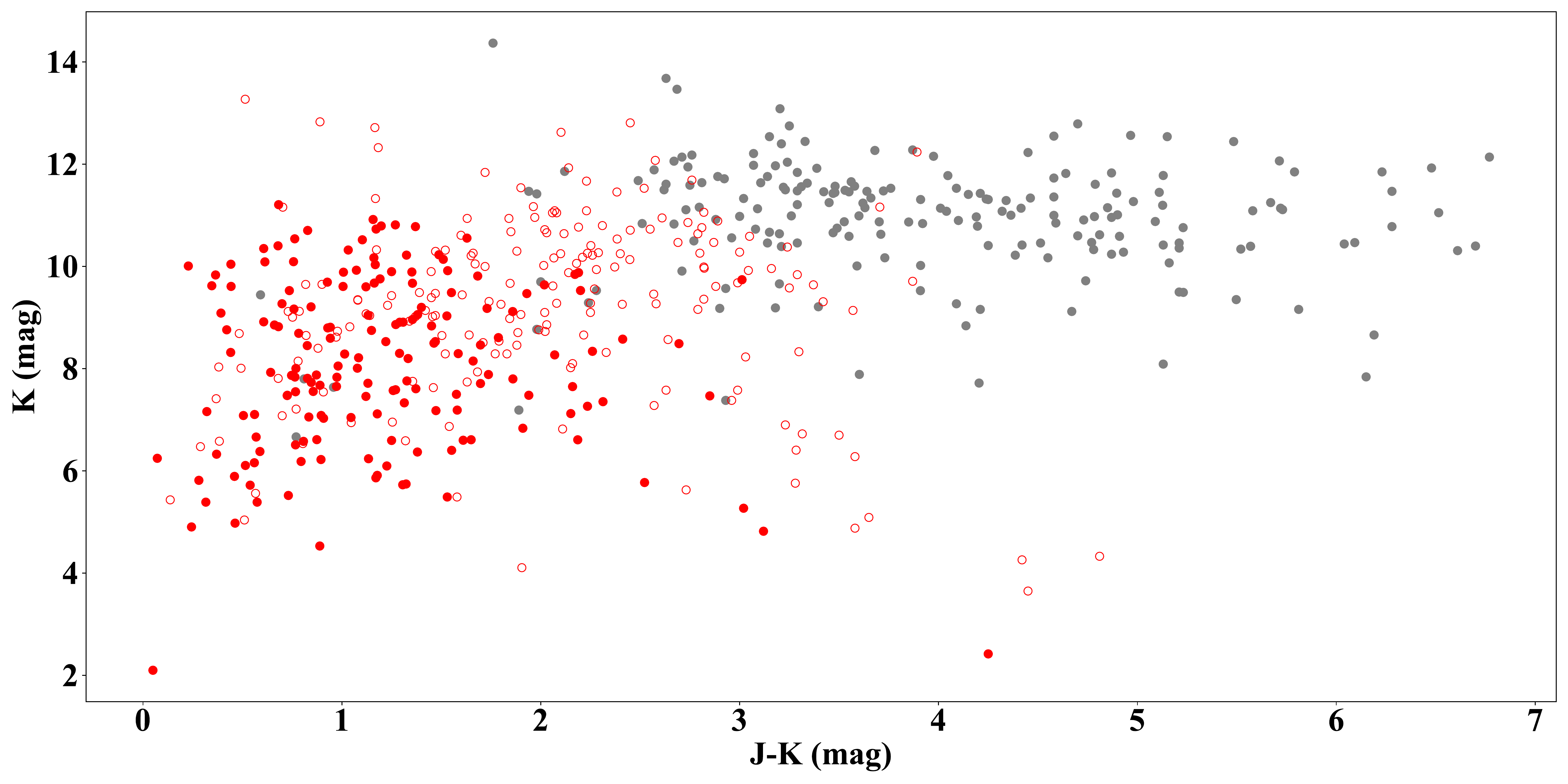} 
  \caption{Plot showing the colour magnitude diagram of Galactic WR stars from the catalogue detected by \textit{Gaia} (red) and WR  stars only observed at IR wavelengths (grey). Stars not observed by \textit{Gaia} have larger (>3) J$-$K colours, indicating significant extinction. Filled red circles are stars with the most reliable distances, these are limited to bright sources (K<12) with J$-$K<3.} 
    \label{fig:K_vs_J}
\end{figure*} 

Wolf-Rayet (WR) stars are the final stages of evolution for massive O stars ($>$25 \Msolar,  \citealt{2007ARA&A..45..177C}). With extremely fast and dense stellar winds, they play an important role in helping to ionize \hii regions and disperse natal gas left over from the star formation process. This feedback may drive and quench star formation. Additionally, WR stars are potential progenitors of long Gamma Ray Bursts \citep{2010A&A...518A..29L} and stripped envelope supernovae, although some may collapse directly to black holes \citep{2009A&A...502..611G}.

The later stages of massive star evolution depend heavily on parameters such as initial mass and metallicity, which influence mass loss rates \citep{2005A&A...429..581M}. Such dependencies make modelling massive star evolution challenging. The accuracy of evolutionary models can be tested with observations, which in turn depend on reliable distances. Inaccurate distances can thus lead to an incorrect understanding of massive star evolution. 

The Milky Way contains a rich population of WR stars, whose total has been estimated at 1200$\pm$200 \citep{2015MNRAS.449.2436R}. Over half have been detected thus far\footnote{\url{http://pacrowther.staff.shef.ac.uk/WRcat/index.php}, v1.21}\addtocounter{footnote}{-1}\addtocounter{Hfootnote}{-1}. Of those, approximately half have been discovered via IR surveys (e.g \citealt{2006MNRAS.372.1407C}, \citealt{2007MNRAS.376..248H}, \citealt{2009AJ....138..402S}), whilst the rest are optically visible. Until now, distances to WR stars have relied upon the small subset of the population, which are  thought to be members of clusters or associations (e.g \citealt{1984A&AS...58..163L}). These stars, along with the WR population of the Magellanic Clouds (e.g \citealt{1968MNRAS.140..409S} and \citealt{1990ApJS...73..685V}),  have been used to calculate absolute magnitude calibrations (e.g \citealt{2001NewAR..45..135V}, \citealt{2015MNRAS.447.2322R}). The calibrations were then applied to estimate distances to field stars. As there is some variation in absolute magnitudes within spectral subtypes, the resulting distances had large uncertainties (50\% according to \citealt{2001NewAR..45..135V}).

Binarity is a key additional piece of the evolutionary puzzle for massive stars. \citet{2009AJ....137.3358M} estimates that 40-70\% of all massive stars are in binaries. Additionally, \citet{2012Sci...337..444S} suggests that 70\% of O stars will undergo interaction during their lifetimes. WR stars may form via Roche Lobe overflow \citep{1967ZA.....65..251K} at the upper end of the stripped star regime \citep{2018A&A...615A..78G} and may be responsible for the high rate of observed Ibc supernovae, relative to the number of massive stars (\citealt{2013MNRAS.436..774E}, \citealt{2011MNRAS.412.1522S}).

Binaries therefore have a major influence on the evolutionary trajectory of massive stars. Studying the fractions of runaways can provide an insight into how massive binaries interact and verify models involving binary physics. Here, again, accurate distances are essential to determine how far a WR star has travelled over its lifetime. 

The second \textit{Gaia} data release (\citealt{2018AA...616A...1G}, \citealt{2016A&A...595A...1G}, hereafter referred to as DR2) offers parallaxes, proper motions and positions for over a billion stars in the Galaxy. A large fraction of the Galactic WR population have been detected in the \textit{Gaia} G band, (330-1050nm) and so \textit{Gaia} increases the number of WR with trigonometric parallaxes from just one (WR11 in Hipparcos, \citealt{2007A&A...474..653V}) to almost 400.

In this work (Paper I) we present distances obtained using \textit{Gaia} data and discuss the resulting new insights into Wolf-Rayet absolute magnitudes, runaways and physical parameters. In Section ~\ref{sec:dist}, we determine the most likely distances for Galactic WR stars using a Bayesian method and in Section ~\ref{sec:absmag}, validate these using absolute magnitudes. We compare the new \textit{Gaia} distances to previous values in Section ~\ref{sec:distdisc}. Distances from the Galactic midplane are discussed in section ~\ref{sec:hab} and used to identify potential runaways. Finally, we conclude with an overview and anticipate potential improvements from later \textit{Gaia} data releases.

In Paper II (Rate, Crowther \& Parker, submitted), we will use these new distances and other Gaia DR2 results to reevaluate WR membership of clusters and associations, and discuss the implications of the results on our understanding of massive star origins and evolution. Future studies will use our distances and extinctions to calculate updated WR line luminosity calibrations for application to unresolved extragalactic WR populations.


\section{Distance determination methods}\label{sec:dist}


\subsection{\textit{Gaia} DR2 catalogue} \label{ssec:gcat}

The parallax and errors used to calculate distances were taken from the \textit{Gaia} DR2 catalogue \citep{2018AA...616A...1G}. The calculation also made use of $G$ band magnitudes, astrometric excess noise (to identify potentially spurious results) and \textit{Gaia} RA and Declination coordinates.

A python {\scriptsize{ASTROQUERY}} (\citealt{2013A&A...558A..33A}, \citealt{2018AJ....156..123A}) script downloaded data from the \textit{Gaia} archive \citep{2017A&C....21...22S} using the ADQL query in Appendix A of the online material. The script searched for stars which were within 1'' of the quoted WR coordinates. Almost all known WR stars are isolated enough for this constraint to be sufficient. The majority (370) of 415 successful search coordinates came from \citet{2001NewAR..45..135V}. However, 45 coordinates from the catalogue did not lead to correct \textit{Gaia} detections. In these instances, coordinates from {\scriptsize{SIMBAD}} were used instead (\citealt{2000A&AS..143....9W}, accessed on 23/05/2018). We checked the coordinates for accuracy using images from {\scriptsize{VPHAS+}} DR3 \citep{2014MNRAS.440.2036D}, {\scriptsize{IPHAS}} DR2 (\citealt{2014MNRAS.444.3230B}, \citealt{2005MNRAS.362..753D}) and 2MASS \citep{2006AJ....131.1163S}, to ensure they corresponded to isolated WR stars. The remaining 243 WR stars yielded no successful results with either coordinate set. Figure ~\ref{fig:K_vs_J} shows most of these (>230) have J--K $>$ 3 mag, indicating
significant foreground dust extinction and are therefore inaccessible to Gaia. 

\begin{figure}
	\centering
	\includegraphics[width=\linewidth]{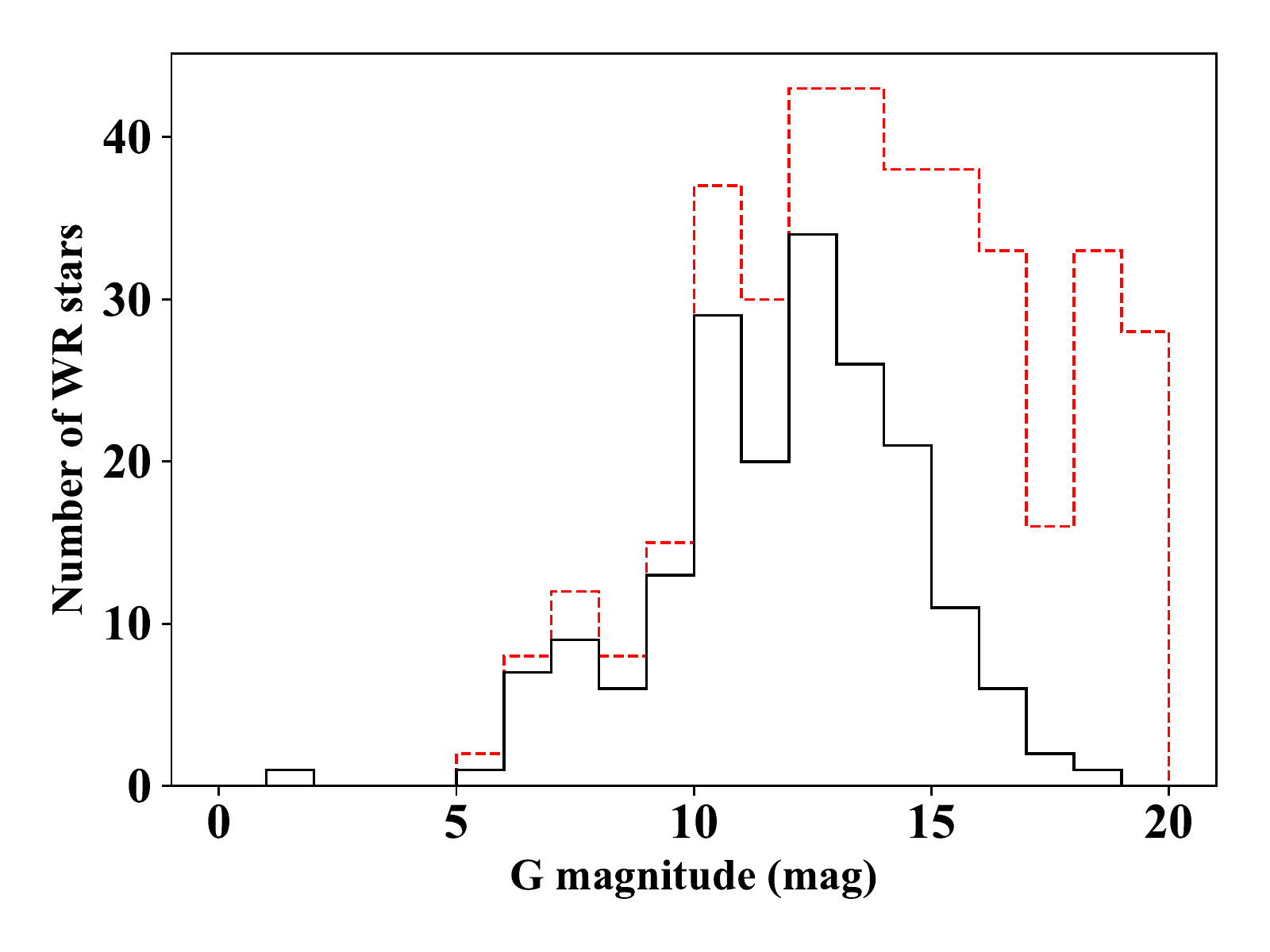}
	\caption{Histogram of $G$ band magnitudes for Gaia DR2 detected WR stars. The solid line (black) involves 187 WR stars with reliable absolute magnitudes (Section \ref{sec:absmag})  and the dashed line (red) involves the full sample of 383 WR stars.} 
    \label{fig:g_hist}
    \end{figure} 

383 stars (58\% of the total) from the Galactic WR catalogue\footnotemark have \textit{Gaia} parallaxes. Of those, 305 have positive parallaxes. Figure ~\ref{fig:g_hist} shows that both the total WR population, and the sample containing only the results with reliable distances, appear to be relatively complete up to $G\sim$13 mag. However, for results with robust absolute magnitudes, the distribution falls off more quickly beyond $G\sim$13 mag. This is because fainter magnitudes are preferentially removed due to their larger astrometric excess noise and increased incidence of negative parallaxes (which are more likely to produce unacceptable absolute magnitudes).


\subsection{Bayesian methods} \label{sssec:bmeth}

The conversion of \textit{Gaia} parallaxes to distances significantly modifies the shape of the original parallax ($\omega$) probability distribution, which means uncertainties do not transform symmetrically. This occurs unless the parallax errors ($\sigma_\omega$) are very small ($\sigma_\omega/\omega<0.1$, \citealt{2015PASP..127..994B}), which is not the case for most of our DR2 sources Additionally, many sources have negative parallaxes; a consequence of the data processing algorithm fitting noisy observations \citep{2018A&A...616A...9L} and of the variation in parallax zero points (see Section \ref{sssec:lhood}). Obtaining the WR star distances should therefore be done carefully using Bayesian methods.

Bayesian inference is therefore the recommended way to transform parallaxes to distances \citep{2018A&A...616A...9L}. The end result is a probability distribution with correct uncertainties, reflecting the non symmetric transformation of parallax to distance. Bayesian methods are also capable of elegantly accounting for unphysical parallaxes and so there is no need to cut negative data from the sample \citep{2018A&A...616A...9L}. 

The technical details of the Bayesian method used, including equations and plots of the model \hii region and dust maps, are in Appendices B, C and D in the online material.       

\begin{figure}
    \centering
    \includegraphics[width=\linewidth]{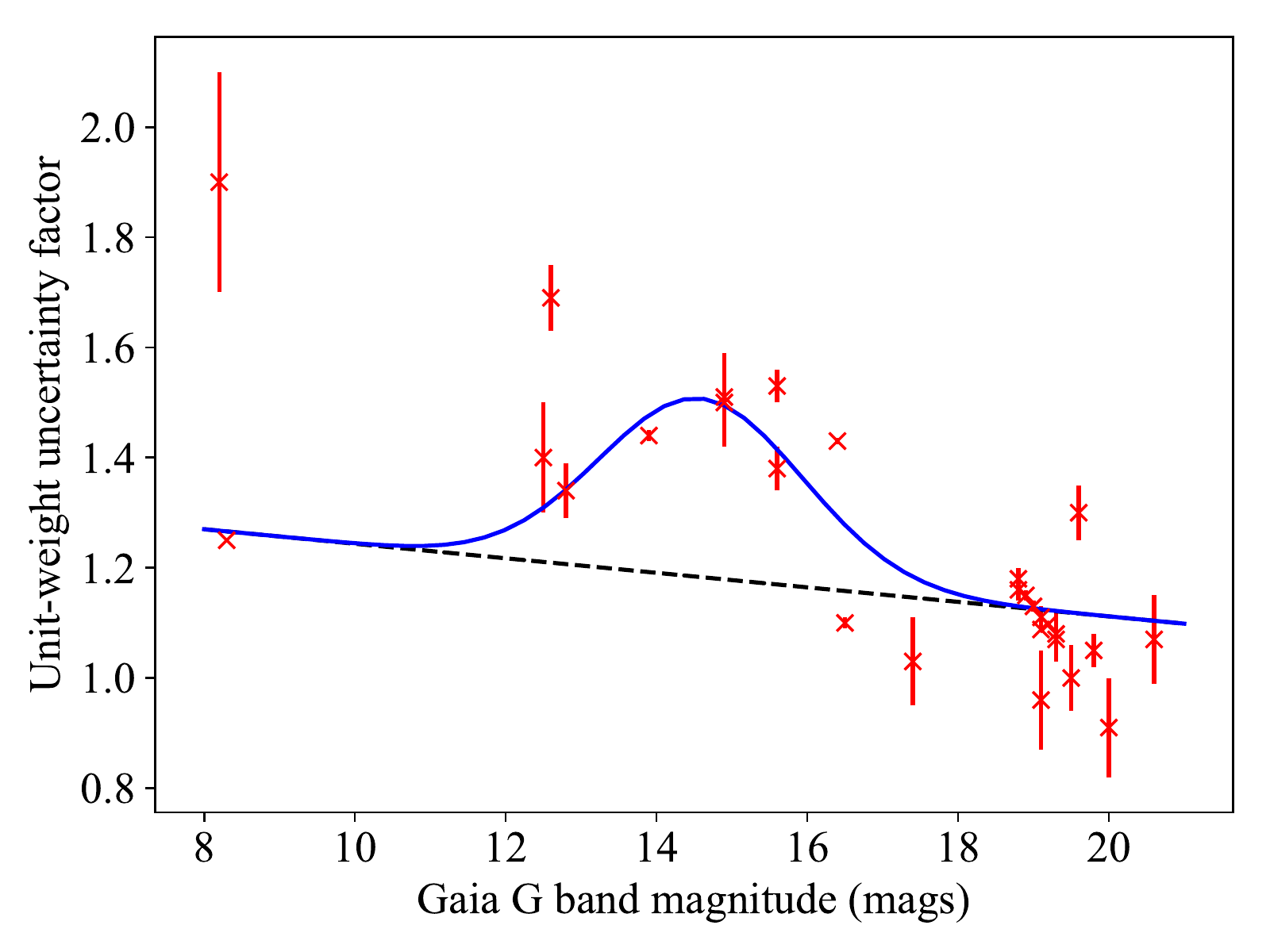} 
    \caption{Weighted fit to the unit weight uncertainty factors from \citet{2018A&A...616A..17A}, used to increase the uncertainties $\sigma_\omega$, to account for underestimation in the \textit{Gaia} catalogue. The dotted line is the linear component of the fit, whilst the solid line is the total fit and the red crosses are the unit weight uncertainties of the external data.} 
    \label{fig:uwu}
\end{figure}

\subsubsection{Likelihoods} \label{sssec:lhood}

The likelihood can be constructed by assuming the parallax distribution is Gaussian, with a mean at the parallax measured by \textit{Gaia} and the parallax error as the standard deviation (\citealt{2018arXiv180407766H}, \citealt{2018A&A...616A...9L}, \citealt{2015PASP..127..994B}). 

The parallaxes quoted by \textit{Gaia} are not corrected for the global zero point. As our sample of WR stars is spread over the sky and the zero point will therefore not be dominated by regional systematics, we choose to apply this global correction to the distance calculation \citep{2018A&A...616A..17A}. In light of the variation in measured zero points and the fact that \citet{gaia_pres} states that the zero point is likely multivariate, with no general process currently available to calculate it, we choose to use the globally measured QSO zero point of $-$0.029 mas (\citealt{2018A&A...616A...2L}, \citealt{2018A&A...616A...9L}). One possible effect of this on the final distances is that if the full multivariate zero point could be used, some small negative parallaxes could be converted to positive values. We discuss further effects of this choice in Section \ref{ssec:compold}.

Additionally, analysis from \citet{2018A&A...616A..17A} suggests that, when compared to external data, the errors of DR2 parallaxes in the catalogue are underestimated. This is because they are consistent with the internal uncertainties, and do not account for  systematics. The underestimation varies with G band magnitude and is particularly acute for results in the range 12<G<15, which could be underestimated by 30-50\% \citep{2018AA...616A...1G}. 

To account for this, we calibrate the uncertainties of Gaia parallaxes using parallaxes from previous surveys. \citet{2018A&A...616A..17A} provide in their Table 1 the unit weight error calculated using a variety of comparative surveys and the median G band of these surveys. Using this data, we present the conversion curve shown in Figure~\ref{fig:uwu}. This is similar to the approach of \citet{gaia_pres}, although our model neglects the HST measurement (1.9 unit weight error at G=8 mag). It is possible to fit a combined Gaussian and straight line which can increase the size of the uncertainties in proportion to the G band magnitude. Details of the equation used for this fit and the impact of increasing the uncertainties on the distances are in Appendices B and E in the online material.

These increased uncertainties were applied to our WR parallaxes and lead to a likelihood that is appropriate for the WR population.

\subsubsection{Prior} \label{sssec:prior}

The prior is a probability distribution of the expected distances for a given WR star. Previous work with \textit{Gaia} \citep{2018AJ....156...58B} has opted for a smooth, exponentially decreasing prior, with a single parameter that can be tuned based on galactic latitude and longitude. This is designed to follow the distribution of all observed stars within the Milky Way and to provide a distance derived purely from a geometric model. 

Almost all WR stars are found at large (kiloparsec) distances and lie preferentially in the Galactic plane, so their observed distribution will be significantly affected by extinction. Previous priors do not properly account for this, which could be problematic for our sample. 

Instead, we build a prior using \hii regions and a dust model for extinction. \hii regions approximate the spatial distribution of massive stars. They are independent of previous WR distribution maps, avoiding any bias from previous incorrect results and are well sampled across the galaxy (as they are detectable at a broad range of wavelengths).  

\begin{figure}
	\centering
	\includegraphics[width=\linewidth]{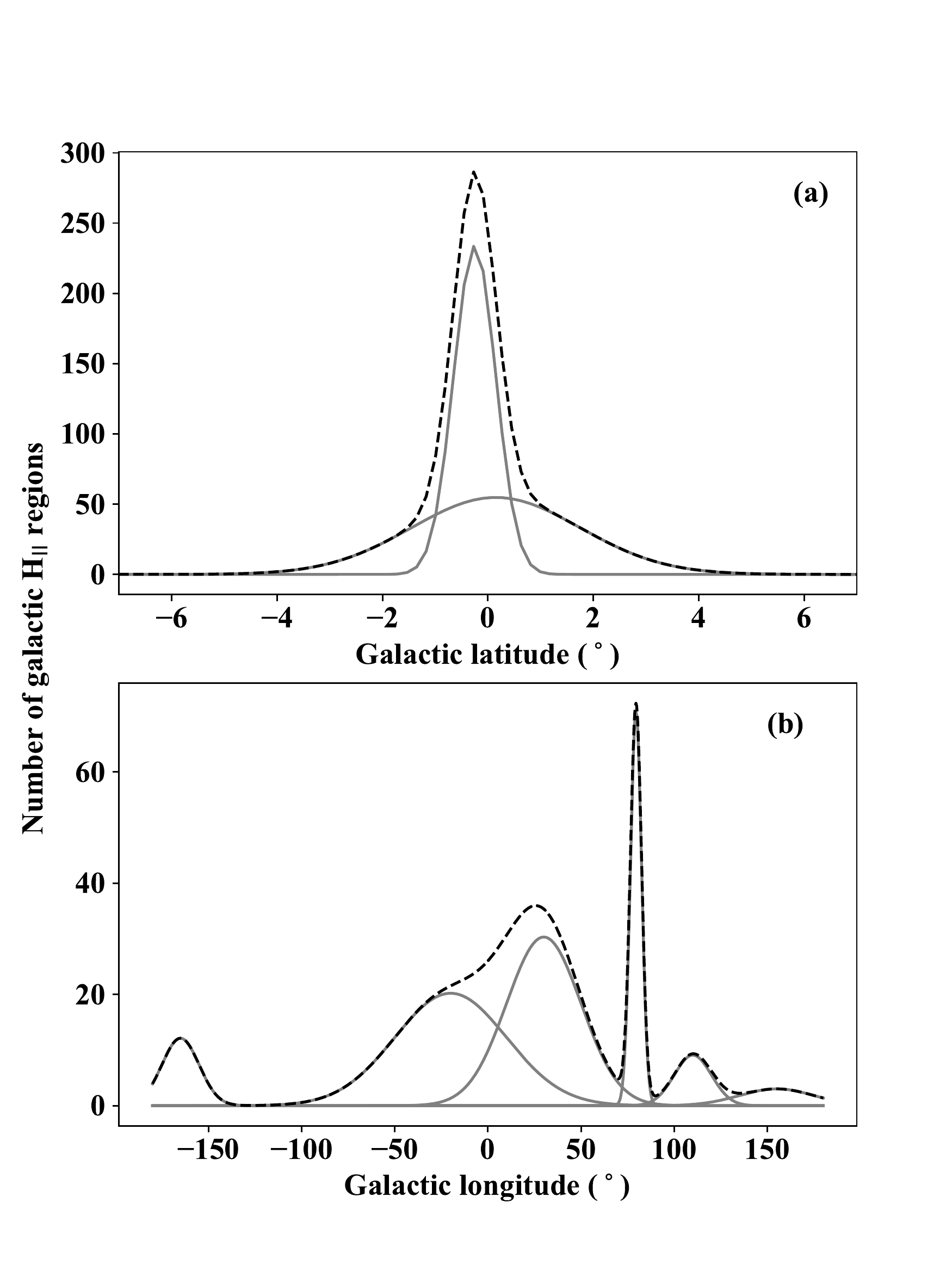} 
  \caption{A mixture of Gaussians showing the number of \hii regions over (a) Galactic latitude and (b) Galactic longitude, based on Figure 6 and data from \citet{2003A&A...397..213P}. The solid lines are the individual Gaussians and the black dotted line is the overall fit. The peak around l=75-90$^{\circ}$ is the Cygnus X region.} 
    \label{fig:mix_gauss}
\end{figure}

To find the overall distribution, we considered \hii region density along each line of sight. Figure ~\ref{fig:mix_gauss} shows a mixture of Gaussians fitted to binned Galactic latitude and longitude distributions, which gave normalised numbers of \hii regions at a given latitude or longitude coordinate. These were then multiplied together to get a total number density  along the line of sight.

We apply a simple dust model \citep{2015MNRAS.447.2322R} to account for the effects of extinction. This consists of both molecular and atomic gas, to replicate the thin and thick disks. For the Sun, we chose a distance of 8.122 kpc \citep{2018A&A...615L..15G} to the Galactic Centre and a height of 20.8 pc \citep{2019MNRAS.482.1417B} above the plane. The resulting distribution is shown in the online supplementary material, in Appendix C.

The prior covered distances between 0 and 15 kpc, at a resolution of 1 pc. The probability is zero below 300 pc, as we do not expect to find any WR stars detected with Gaia closer than this distance. The final form of the prior therefore varies from Gaussian like in regions with a pronounced \hii region peak or low extinction, to exponential like in regions with a less pronounced peak or high extinction.

\begin{figure}
	\includegraphics[width=1.1\linewidth]{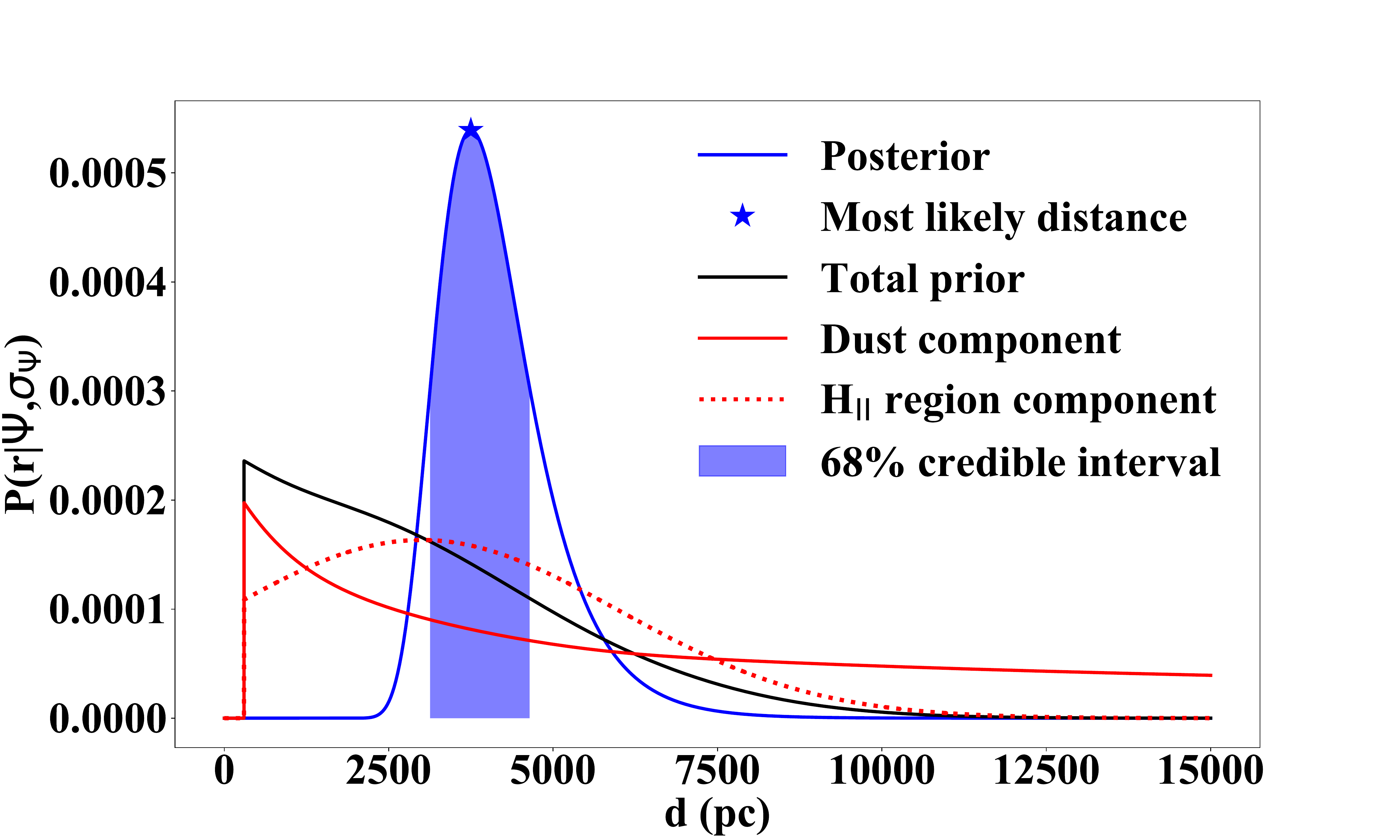} 
  \caption{Posterior distribution for WR4, shown alongside the prior components and credible interval. The filled star is the most likely distance to WR4 (3.75$^{+0.89}_{-0.62}$ kpc, compared to 3.71$^{+0.65}_{-0.49}$ kpc from \citealt{2018AJ....156...58B}).} 
    \label{fig:eg_distrib}
\end{figure}

\subsubsection{Posterior} \label{sssec:post}

We then calculated the posterior distribution. Figure ~\ref{fig:eg_distrib} shows an example of this for WR4, together with the prior and its components. 

Use of the numerical dust model meant we could not differentiate the posterior and produce an analytical solution for the maximum likelihood. Instead the peak of the distribution was taken as the most likely distance. Credible intervals (similar to those used in \citealt{2018AJ....156...58B}) give distances which, when used as integral limits, cover 68\% of the area below the curve. The one sigma errors are the differences between the peak and these distances. 

\begin{figure}
  \includegraphics[width=\linewidth]{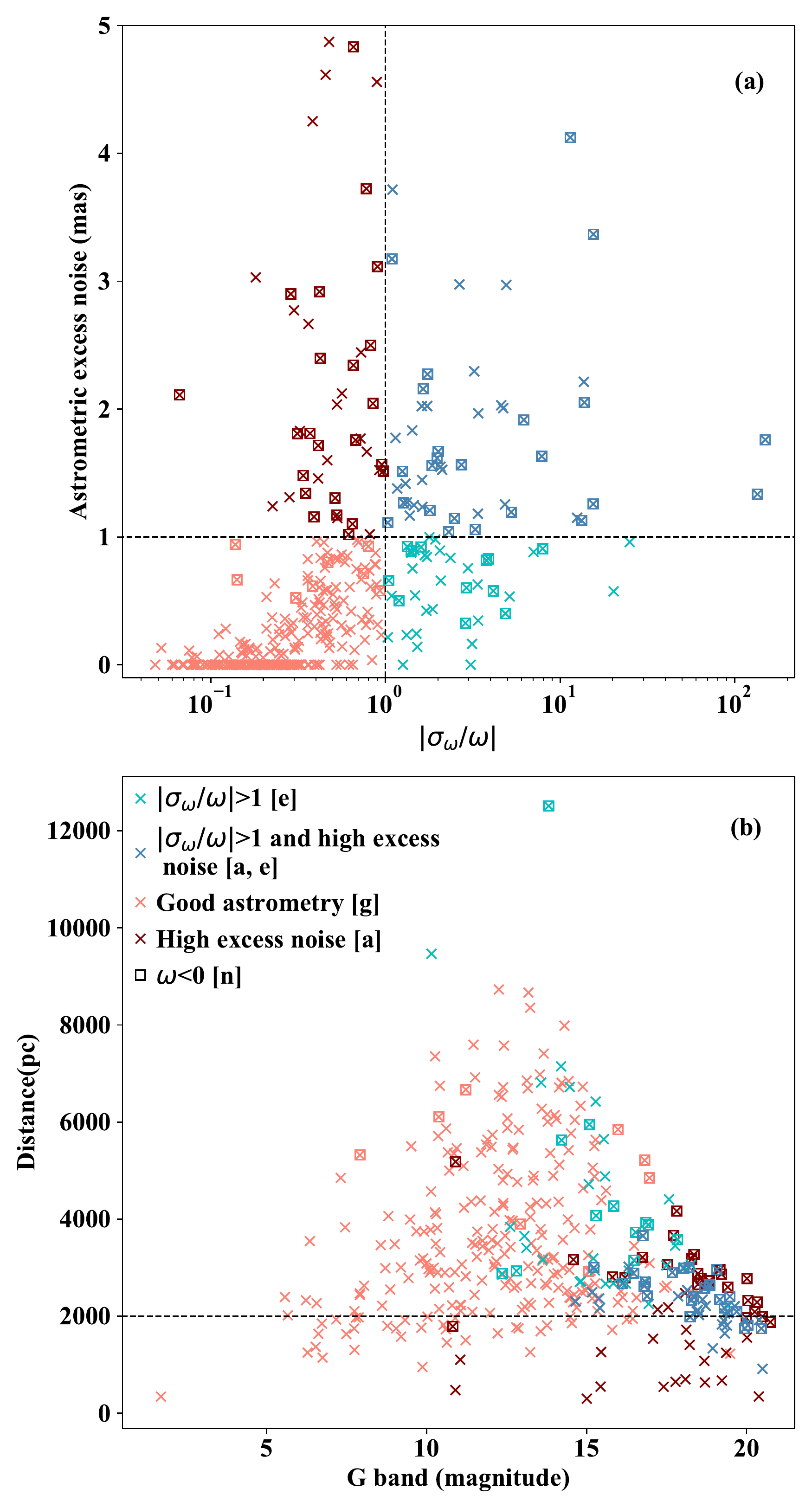}
  \caption{(a) Comparison between parallax error $|\sigma_\omega/\omega|$ and astrometric error noise (mas) for Galactic WR stars from \textit{Gaia} DR2, for which dotted lines indicate values of unity for each parameter to highlight data quality flags a, e, g, n; (b) Comparison between G band magnitudes and inferred distances (pc) for Galactic WR stars from \textit{Gaia} DR2, with the dotted line marking a distance of 2 kpc.}
  \label{fig:err_plt}
\end{figure}

\begin{table*}
  \renewcommand{\arraystretch}{1.5}
  \caption{Gaia DR2 astrometric, photometric and parallax properties for 383 Galactic WR stars, including WR11 using a parallax and photometry from Hipparcos \citep{2007A&A...474..653V}. The distance for WR11 was calculated in the same manner as WR with \textit{Gaia} results, except the adjustments to calculate $\omega$ and $\sigma_{\omega}$ were not applied. Stellar luminosities, updated from \citet{2019A&A...625A..57H} and \citet{2019A&A...621A..92S} according to our revised distances, are restricted to sources with no error flags. The full table is available in the online supplementary material.}
  \begin{tabular}{
  l@{\hspace{2mm}}
  l@{\hspace{3mm}}
  l@{\hspace{3mm}}
 c@{\hspace{3mm}}
  c@{\hspace{3mm}}
  c@{\hspace{3mm}}
  c@{\hspace{3mm}}
  c@{\hspace{3mm}}
  r@{\hspace{2mm}}
 c@{\hspace{2mm}}
  c@{\hspace{3mm}}
  c@{\hspace{3mm}}
  l}
  \hline
WR & Spectral & Alias & RA & Dec & $\omega \pm \sigma_w$ & $d$ & |z| & $G$ & $G_{BP}-G_{RP}$ & Excess & $\log L$ & Flags \\
Number & Type &  & J2015 & J2015 & mas & kpc & pc & mag & mag & Noise & $L_{\odot}$ & \\
  \hline
  WR1 & WN4b & HD 4004 & 00 43 28.39 & +64 45 35.4 & 0.314$\pm$0.040 & 3.15$\substack{+0.47 \\ -0.36}$ & 125$\substack{+15 \\ -12}$ & 9.79 & 1.05 & 0.00 &  & g \\
	WR3 & WN3ha & HD 9974 & 01 38 55.62 & +58 09 22.6 & 0.342$\pm$0.051 & 2.90$\substack{+0.52 \\ -0.39}$ & 188$\substack{+37 \\ -27}$ & 10.58 & 0.18 & 0.10 & 5.56 & g \\
	WR4 & WC5+? & HD 16523 & 02 41 11.67 & +56 43 49.8 & 0.258$\pm$0.051 & 3.75$\substack{+0.89 \\ -0.62}$ & 174$\substack{+46 \\ -32}$ & 9.68 & 0.51 & 0.06 & 5.72 & g \\
	WR5 & WC6 & HD 17638 & 02 52 11.66 & +56 56 07.1 & 0.334$\pm$0.042 & 2.97$\substack{+0.43 \\ -0.33}$ & 90$\substack{+16 \\ -12}$ & 10.06 & 0.94 & 0.00 & 5.53 & g \\
	WR6 & WN4b & EZ CMa & 06 54 13.04 & $-$23 55 42.0 & 0.441$\pm$0.065 & 2.27$\substack{+0.42 \\ -0.31}$ & 376$\substack{+73 \\ -53}$ & 6.57 & 0.04 & 0.18 & 5.78 & g \\
	WR7 & WN4b & HD 56925 & 07 18 29.13 & $-$13 13 01.5 & 0.221$\pm$0.051 & 4.23$\substack{+1.08 \\ -0.74}$ & 11$\substack{+2 \\ -1}$ & 11.17 & 0.73 & 0.00 & 5.33 & g \\
	WR8 & WN7o/CE & HD 62910 & 07 44 58.22 & $-$31 54 29.5 & 0.263$\pm$0.038 & 3.74$\substack{+0.63 \\ -0.48}$ & 226$\substack{+41 \\ -31}$ & 9.92 & 0.84 & 0.00 &  & g \\
	WR9 & WC5+O7 & HD 63099 & 07 45 50.40 & $-$34 19 48.5 & 0.212$\pm$0.035 & 4.57$\substack{+0.84 \\ -0.63}$ & 256$\substack{+70 \\ -52}$ & 10.14 & 1.30 & 0.00 &  & g \\
	WR10 & WN5h  & HD 65865 & 07 59 46.24 & $-$28 44 03.0 & 0.162$\pm$0.040 & 5.46$\substack{+1.25 \\ -0.91}$ & 75$\substack{+12 \\ -9}$ & 10.94 & 0.60 & 0.09 &5.78 & g \\
	WR11 & WC8+O7.5III-V & $\gamma$ Vel & 08 09 31.96 & $-$47 20 11.8 & 2.920$\pm$0.300 & 0.34$\substack{+0.04 \\ -0.03}$ & 24$\substack{+5 \\ -4}$ & 1.70 &   &  &  &  \\
  WR12 & WN8h & Ve5-5 & 08 44 47.29 & $-$45 58 55.4 & 0.154$\pm$0.037 & 5.71$\substack{+1.24 \\ -0.92}$ & 175$\substack{+42 \\ -31}$ & 10.36 & 1.15 & 0.00 & 5.93 & g \\
  \hline
  \end{tabular}
  \label{table:final}
Columns are: (1) WR Number, (2) Spectral type, (3) Alternative name, (4) \textit{Gaia} Right Ascension, (5) \textit{Gaia} Declination, (6) Zero point corrected parallax $\omega$ and inflated error $\sigma_{\omega}$, (7) Distance from the Sun, (8) Distance from the midplane, (9) \textit{Gaia} G band apparent magnitude, (10)  \textit{Gaia} colour index, (11) Astrometric excess noise, (12) Stellar luminosity, (13) Error flags, a = astrometric excess noise $>$ 1 mas; e = large parallax uncertainty $|\sigma_\omega/\omega|$>1; n = negative parallax $\omega$<0, g = good astrometry.
\end{table*}

\begin{table*}
  \renewcommand{\arraystretch}{1.5}
  \caption{Intrinsic colours of WR stars from PoWR models (\citealt{2004A&A...427..697H}
      and \citealt{2015A&A...579A..75T} for WN, \citealt{2012A&A...540A.144S} for WC)
    for $\mathrm{(b-v)_0^{WR}}$ and monochromatic (J$-$K)$^{\rm mono}_{0}$ and
    (H$-$K)$^{\rm mono}_{0}$, and \citet{2015MNRAS.447.2322R} for
    (J$-\mathrm{K_s}$)$_0$ and (H$-\mathrm{K_s}$)$_0$.}
   \begin{tabular}{
  l@{\hspace{2mm}}
  l@{\hspace{3mm}}
  c@{\hspace{3mm}}
 c@{\hspace{3mm}}
  c@{\hspace{3mm}}
  c@{\hspace{3mm}}
  c@{\hspace{3mm}}
  c@{\hspace{3mm}}
  c}
  \hline
  WR subtype & PoWR model & $\log( T/k)$ & $\mathrm{\log(R_t)}$ & $(\mathrm{b-v})_0^{WR}$ & (J$-\mathrm{K_s}$)$_0$ & (H$-\mathrm{K_s}$)$_0$ & (J$-$K)$^{\rm mono}_{0}$ & (H$-$K)$^{\rm mono}_{0}$\\
  \hline
  WN3-4 & WNE 12-11 & 4.95 & 1.0 & $-$0.32$\pm$0.1 & $-$0.11$\pm$0.1 & $-$0.03$\pm$0.1 & \phantom{$-$}0.24 &  \phantom{$-$}0.16 \\
  WN4b-7b & WNE 12-18 & 4.95 & 0.3 & $-$0.18$\pm$0.1 & \phantom{$-$}0.37$\pm$0.1 & \phantom{$-$}0.27$\pm$0.1 &  \phantom{$-$}0.63 &  \phantom{$-$}0.40 \\
  WN5-6 & WNE 08-11 & 4.75 & 1.0 & $-$0.28$\pm$0.1 & \phantom{$-$}0.18$\pm$0.1 & \phantom{$-$}0.16$\pm$0.1 &  \phantom{$-$}0.30 &  \phantom{$-$}0.20 \\
  WN7-9 & WNL 06-13 & 4.65 & 0.8 & $-$0.15$\pm$0.1 & \phantom{$-$}0.13$\pm$0.1 & \phantom{$-$}0.11$\pm$0.1 &  \phantom{$-$}0.30 &  \phantom{$-$}0.18 \\
  WN6ha & WNL 07-07 & 4.70 & 1.4 & $-$0.33$\pm$0.1 & $-$0.015$\pm$0.1 & \phantom{$-$}0.03$\pm$0.1 &  \phantom{$-$}0.00 &  \phantom{$-$}0.00 \\
  WN7ha & WNL 07-07 & 4.70 & 1.4 & $-$0.33$\pm$0.1 & $-$0.04$\pm$0.1 & \phantom{$-$}0.01$\pm$0.1 &  \phantom{$-$}0.00 &  \phantom{$-$}0.00 \\
  WN8-9ha & WNL 05-07 & 4.60 & 1.4 & $-$0.32$\pm$0.1 & $-$0.04$\pm$0.1 & \phantom{$-$}0.01$\pm$0.1 &  \phantom{$-$}0.01 &  \phantom{$-$}0.00 \\
  Of/WN & WNL 07-06 & 4.65 & 1.5 & $-$0.34$\pm$0.1 & $-$0.11$\pm$0.1 & $-$0.07$\pm$0.1 & $-$0.04 & $-$0.03 \\
  WO2-3 & WC 17-12 & 5.20 & 0.9 & $-$0.37$\pm$0.1 & \phantom{$-$}0.11$\pm$0.1 & \phantom{$-$}0.00$\pm$0.1 &  \phantom{$-$}0.20 &  \phantom{$-$}0.11 \\
  WC4-7 & WC 11-16 & 4.90 & 0.5 & $-$0.20$\pm$0.2 & \phantom{$-$}0.62$\pm$0.1 & \phantom{$-$}0.58$\pm$0.2 &  \phantom{$-$}0.54 &  \phantom{$-$}0.33 \\
  WC8 & WC 09-14 & 4.80 & 0.7 & $-$0.37$\pm$0.1 & \phantom{$-$}0.43$\pm$0.1 & \phantom{$-$}0.38$\pm$0.1 &  \phantom{$-$}0.38 &  \phantom{$-$}0.21 \\
  WC9 & WC 06-12 & 4.65 & 0.9 & $-$0.32$\pm$0.1 & \phantom{$-$}0.23$\pm$0.1 & \phantom{$-$}0.26$\pm$0.1 &  \phantom{$-$}0.12 &  \phantom{$-$}0.09 \\
  WN/WC &  &  &  & $-$0.23$\pm$0.1 & \phantom{$-$}0.37$\pm$0.1 & \phantom{$-$}0.27$\pm$0.1 &  & \\
  \hline
  \end{tabular}
  \label{table:powr_types}
\end{table*}

\subsection{Flags from \textit{Gaia}} \label{ssec:gflag}

The validity of the distances is determined by the quality of the parallax data. A significantly negative parallax (less than the zero point), will result in a smaller likelihood than a positive parallax and will increase the proportional size of the prior. Negative parallaxes can also indicate unreliable \textit{Gaia} data. Similarly, a large error (on the scale of the data itself) will also result in a much smaller likelihood and a greater influence from the prior. 

These issues mainly arise from badly fitted parallax solutions, which can be identified using parameters in the \textit{Gaia} catalogue. We chose astrometric excess noise (the observational noise which needs to be added to the data to match the solution residuals) as this identifier. Large values can indicate that a solution does not fit the data well. We chose to use this parameter, as it was the quality indicator with the clearest cut-off and acted as a good benchmark for removing bad values when calculating absolute magnitudes. The excess noise can also account for modelling errors, which are not included in the observational noise. Significant astrometric excess noise is mainly applied to fainter objects, in particular those with brighter neighbours. 

The \textit{Gaia} documentation \citep{2018gdr2.reptE..14H} states that high excess noise will be present in early releases and suggests that users apply their own cut-offs to determine erroneous values. The ideal excess for results with distances is zero, which indicates a good fit. However, excluding an outlier with excess noise 18 mas, the average value for our sample is 0.71 mas and the standard deviation is 0.98 mas. Therefore, we flag all results with noise above 1 mas. 

Combined, our three criteria for flagging \textit{Gaia} data quality are

{\fontfamily{qcr}\selectfont 
a = astrometric\_excess\_noise}>1

e = $|\sigma_\omega/\omega|$>1 

n = $\omega$<0.
 
Results without any of these issues are given the 'g' flag. These flags are applied to the distances in Table ~\ref{table:final}.

We apply the flags to the zero point corrected parallaxes and the increased errors, as these are the values are used to calculate distance. A star can be flagged if it satisfies one or more of the criteria. If all three are applied, then ~37\% of the WR stars with parallaxes have an a, e or n flag. 

59\% of the flagged results had more than one negative flag. This reflects the way such errors are intertwined, where a poor solution fit due to noisy observations can lead to a large astrometric excess noise, sizeable errors and negative parallaxes all at once. 

The relations between flags are shown in Figure ~\ref{fig:err_plt}. In general, WR stars with large astrometric excess noise are supposedly located closer than 4 kpc, and in many cases closer than 2 kpc. This latter group further breaks down into brighter objects at around $G$=11 mag (WR146 and WR115) and $G$=15 mag (including WR77p) and fainter objects with $G$>17 mag. The fainter objects may have high excess noise because of astrometric modelling difficulties, caused by issues like binarity or a badly determined spacecraft attitude during a given time interval (\citealt{2018gdr2.reptE..14H}, \citealt{2018A&A...616A...2L}). These problems would make it difficult for the \textit{Gaia} AGIS algorithm to reliably extract astrometric parameters. The brighter objects may have high excess noise for a variety of reasons, such as issues with instrument calibration \citep{2018A&A...616A...2L}. High astrometric excess noise can also occur if the stars are in binaries (WR146) or potential binaries (WR115). 

The other two flags show a less clear breakdown. Negative parallaxes can occur at all magnitudes and distances, but have non zero excess noise. Only a small fraction of results with large error ratios have zero astrometric excess noise and none at all occur below $G$=12 mag. Both flags become increasingly common beyond $G$=15 mag and only a few points beyond $G$=18 mag are not flagged. This is expected given that highly reddened objects at any distance are more difficult for \textit{Gaia} to observe. The flags applied to the data are listed in Table ~\ref{table:final}. Any users should note that distances to these flagged stars may be suspect and should account for this in their analysis. 


\section{Absolute magnitudes} \label{sec:absmag}

In addition to the \textit{Gaia} data quality flags, we checked the validity of the distance results by calculating absolute magnitudes in the  $\mathrm{v^{WR}}$- band \citep{1968MNRAS.140..409S}\footnote{A 'WR' superscript is added to distinguish the Smith $v$ filter  from the standard Johnson V-band filter}, designed to avoid WR emission lines, and the $\mathrm{K_s}$ band. As part of this, we calculated extinction using intrinsic colours and an adopted extinction law. The result was then combined with distances and apparent magnitudes to obtain absolute magnitudes. 

\subsection{Intrinsic colours for single stars} \label{ssec:intcol}

Intrinsic optical colours were taken from PoWR grids (\citealt{2004A&A...427..697H} and \citealt{2015A&A...579A..75T} for WN, \citealt{2012A&A...540A.144S} for WC), for single stars in the $\mathrm{v^{WR}}$ band (see Table ~\ref{table:powr_types}). The exception is for WN/WC stars, as the value 
$\mathrm{(b-v)_{0}^{WR}} = -$0.23 is averaged from the $E(\mathrm{b-v)^{WR}}$ values of \citet{2012A&A...540A.144S} and the $\mathrm{b^{WR}}$ and $\mathrm{v^{WR}}$ apparent magnitudes of each star. Intrinsic colours for the J, H and $\mathrm{K_s}$ bands are taken from \citet{2015MNRAS.447.2322R}, with monochromatic near-IR PoWR synthetic colours also included.

\subsection{Intrinsic colours for binary systems} \label{ssec:bin}

\begin{figure}
	\centering
	\includegraphics[width=\linewidth]{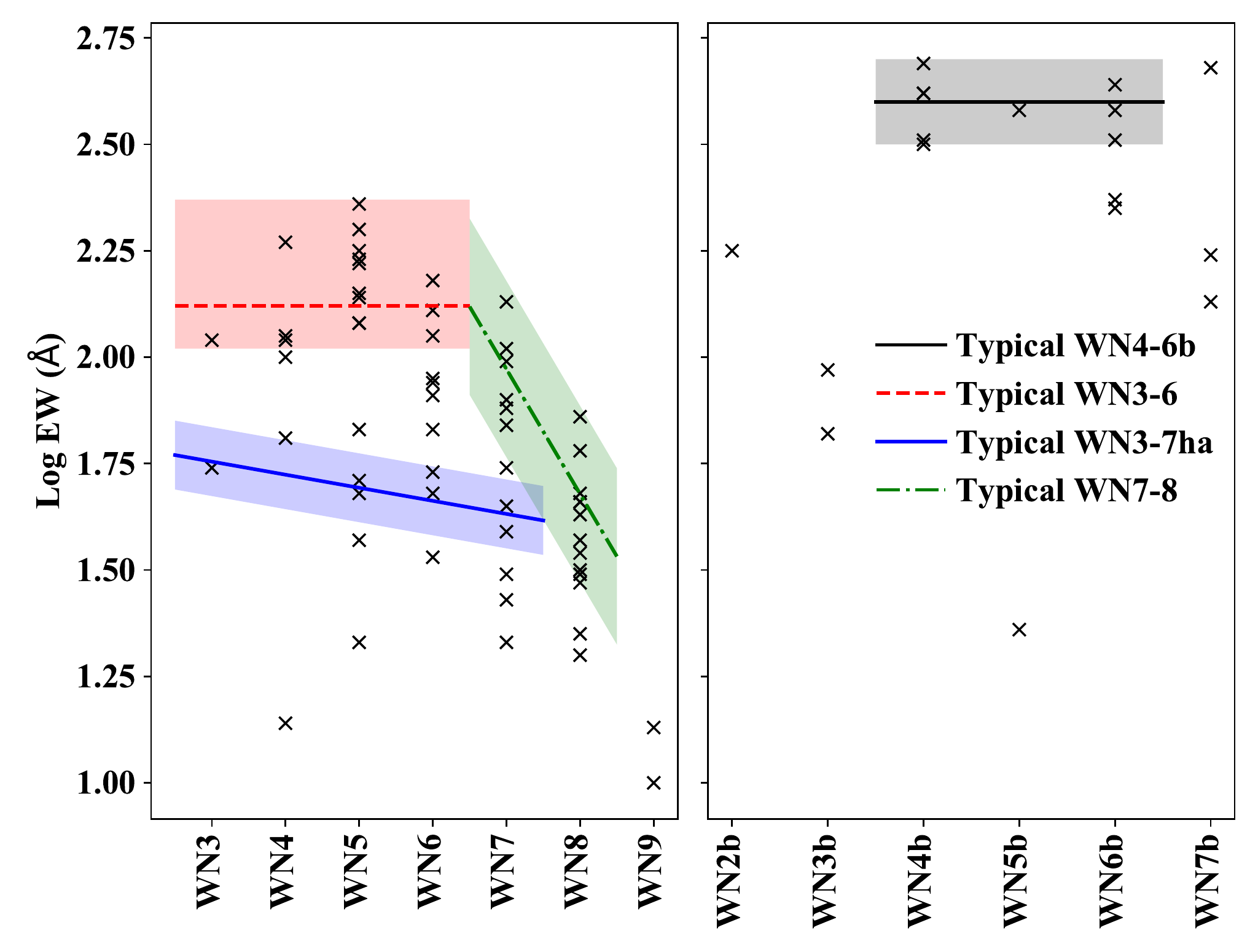} 
	\caption{WN stars with He{\scriptsize{II}} 4686\ang\ equivalent widths from \citet{1989ApJ...337..251C} and \citet{1996MNRAS.281..163S}. The lines show the equivalent width for a typical single WN star at each subtype. The shaded regions should contain only single stars.} 
    \label{fig:wn_lines}
\end{figure}

\begin{figure}
	\includegraphics[width=1.1\linewidth]{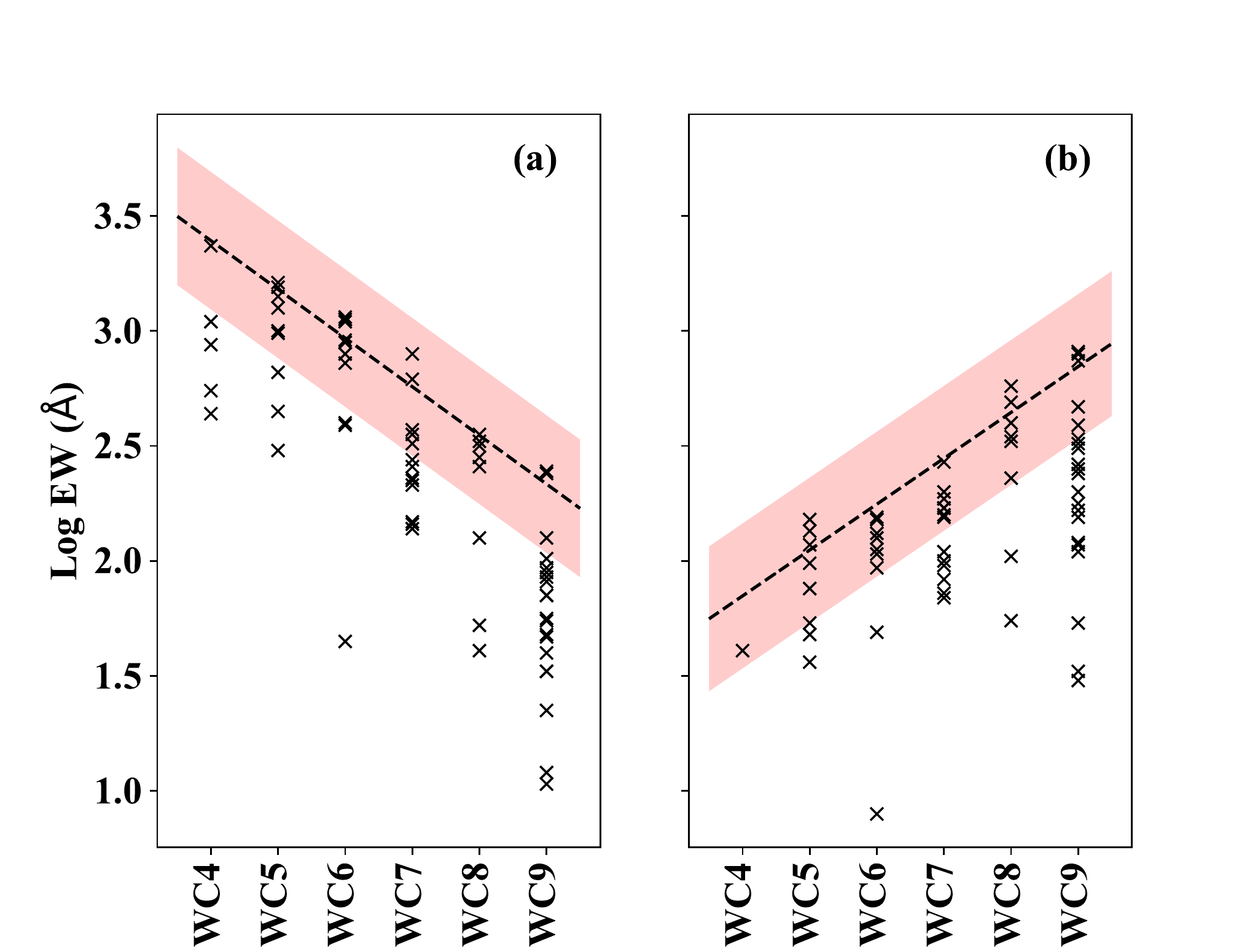} 
	\caption{Equivalent widths of (a) C{\scriptsize{IV}} 5808\ang\  and (b) C{\scriptsize{III}} 5696\ang\ from \citet{1986ApJ...300..379T}, \citet{1989ApJ...337..251C}, \citet{1990ApJ...358..229S}, \citet{1991ApJ...378..302C}, \citet{2009PASP..121..591M} and \citet{2014MNRAS.445.1663Z} showing the relation between line strengths and spectral types for both single and binary stars. The dotted line shows the equivalent width for a typical single WC star at each subtype. The shaded region is the one sigma standard deviation and should contain only single stars.} 
    \label{fig:wc_lines}
\end{figure}

16\% (61 stars) of our WR sample were classified as binaries. For these systems, we calculated absolute magnitudes in the same manner as single stars, but included the companion in the intrinsic colour by measuring the dilution of the strongest optical emission lines. These are
He{\scriptsize{II}} 4686\ang\ for WN stars, and
C{\scriptsize{IV}} 5808\ang\ and C{\scriptsize{III}} 5696\ang\ for WC stars.
We fit the relation of the equivalent width to subtype for single stars (see Figs~\ref{fig:wn_lines}--\ref{fig:wc_lines}),
to obtain the equivalent width of a 'typical' single star with a particular subtype. 

For WC stars, we used C{\scriptsize{IV}} 5808\ang to obtain the typical equivalent width of a single WR star with subtype 4, 5 or 6. In subtypes 8 and 9, the dominant line is instead C{\scriptsize{III}} 5696\ang. The fractions for WC7, which can contain either line, were the average dilution of the two. The fractional contribution of the WR's visible light ($Fc_{WRv}$) to the binary was then found using:
\begin{equation} \label{eq:wr_contrib}
  Fc_{WRv}=\frac{EW_{b}}{EW_{s}}
\end{equation} 
where $EW_{b}$ is the WR equivalent width for the binary and $EW_{s}$ is the equivalent width for a single star. We summed the intrinsic colour of each component, weighted by contribution fraction, to obtain the colour for the system. 

WR stars contribute a higher fraction of the continuum flux to the binary at near-IR wavelengths with respect to the visual (see Table ~\ref{table:sed_ir}). To illustrate this, we compare template spectra from WR stars of different subtypes to an O star from a Kurucz ATLAS model ($T_{\rm eff}$ = 37500K and $\log g$ = 5). Each template spectrum is set to the same V-band continuum flux. The fraction of light contributed by the template O star at IR wavelengths can then be calculated. We use this to obtain the intrinsic colours of the binary in the same way as optical wavelength colours.

\begin{table}
  \caption{The relative continuum flux contribution of WR stars to O-type companions at near-IR wavelengths for various subtypes, adopting a Kurucz ATLAS O star model with $T_{\rm eff}$ = 37500K and $\log g$ = 5 for the companion, assuming each contribute 50\% of the V-band continuum flux.}
  \begin{tabular}{|p{.13\textwidth}|p{.05\textwidth}|p{.05\textwidth}|p{.05\textwidth}|p{.05\textwidth}|}
  \hline
  WR subtypes & & $F_{WR}/F_{0}$ & & \\
  & V & J & H & K \\
  \hline
  WNE-w & 1 & 1.33 & 1.56 & 1.94 \\
  WNE-s & 1 &  2.45 & 3.35 & 4.56 \\
  WN6ha & 1 &  1.22 & 1.38 & 1.63 \\
  WN8 & 1 &  2.03 & 2.70 & 3.55 \\
  WN9 & 1 &  1.33 & 1.5 & 1.78 \\
  Of/WN & 1 & 1.17 & 1.22 & 1.33 \\
  WC4-5 & 1 & 2.03 & 2.57 & 3.55 \\
  WC6-7 & 1 & 1.94 & 2.45 & 3.35 \\
  WC8 & 1 & 1.86 & 2.23 & 3.00 \\
  WC9 & 1 & 1.70 & 2.13 & 2.57 \\
  \hline

  \end{tabular}
  \label{table:sed_ir}
\end{table}

For WR11, we used the light ratio derived in \citet{2000A&A...358..187D} and for WR104, we used the ratio from \citet{2000MNRAS.314...23W}. For WR30a, we estimated the fraction of light contributed by the WR was 10\%, based on the emission line strength of similar WO4 star BAT99-123 (Br93, Sand 2). For WR64-4, we used the He{\scriptsize{II}} 1.16$\mathrm{\mu m}$, 1.69$\mathrm{\mu m}$ and 2.19$\mathrm{\mu m}$ IR lines to find contribution ratios, as no optical data were available. For WR35a, a reverse approach was followed based on
the absolute magnitude of the system and assuming an absolute V magnitude for the O8.5V companion (from \citealt{2006A&A...457..637M}), to calculate the absolute magnitude of the WR component. 

\subsection{Optical and IR extinctions} \label{ssec:extinctions}

We calculate dust extinctions using the intrinsic colours (Table~\ref{table:powr_types}) and apparent magnitudes in the $\mathrm{v^{WR}}$ band taken from the Galactic Wolf-Rayet catalogue, which was primarily compiled from \citet{2001NewAR..45..135V} and \citet{1988AJ.....96.1076T}. J, H and $\mathrm{K_s}$ band magnitudes were primarily sourced from the 2MASS catalogue. The $\mathrm{K_s}$ band extinction, $A_{Ks}$, was calculated using the standard extinction law $A_{Ks} = 0.107 A\substack{WR \\ v}$ (obtained from $A_{Ks} = 0.118 A_V$ from \citealt{1989ApJ...345..245C} and $A\substack{WR \\ v} = 1.1 A_V$ from \citealt{1982IAUS...99...57T}), if values of $A\substack{WR \\ v}$ were available. Otherwise, $A_{Ks}$ was calculated with the relations of $A_H$ and $A_J$ to $A_{Ks}$ (using parameters from \citealt{2011ApJ...737...73F} towards the Galactic Centre and \citealt{2009MNRAS.400..731S} elsewhere, as in \citealt{2015MNRAS.447.2322R}). 

For WR25, known to have an anomalous extinction curve, we calculated $A\substack{WR \\ v}$ using $R\substack{WR \\ v}=6.2$ from \citet{1995A&A...293..427C}.

\begin{figure}
	\centering
	\includegraphics[width=\linewidth]{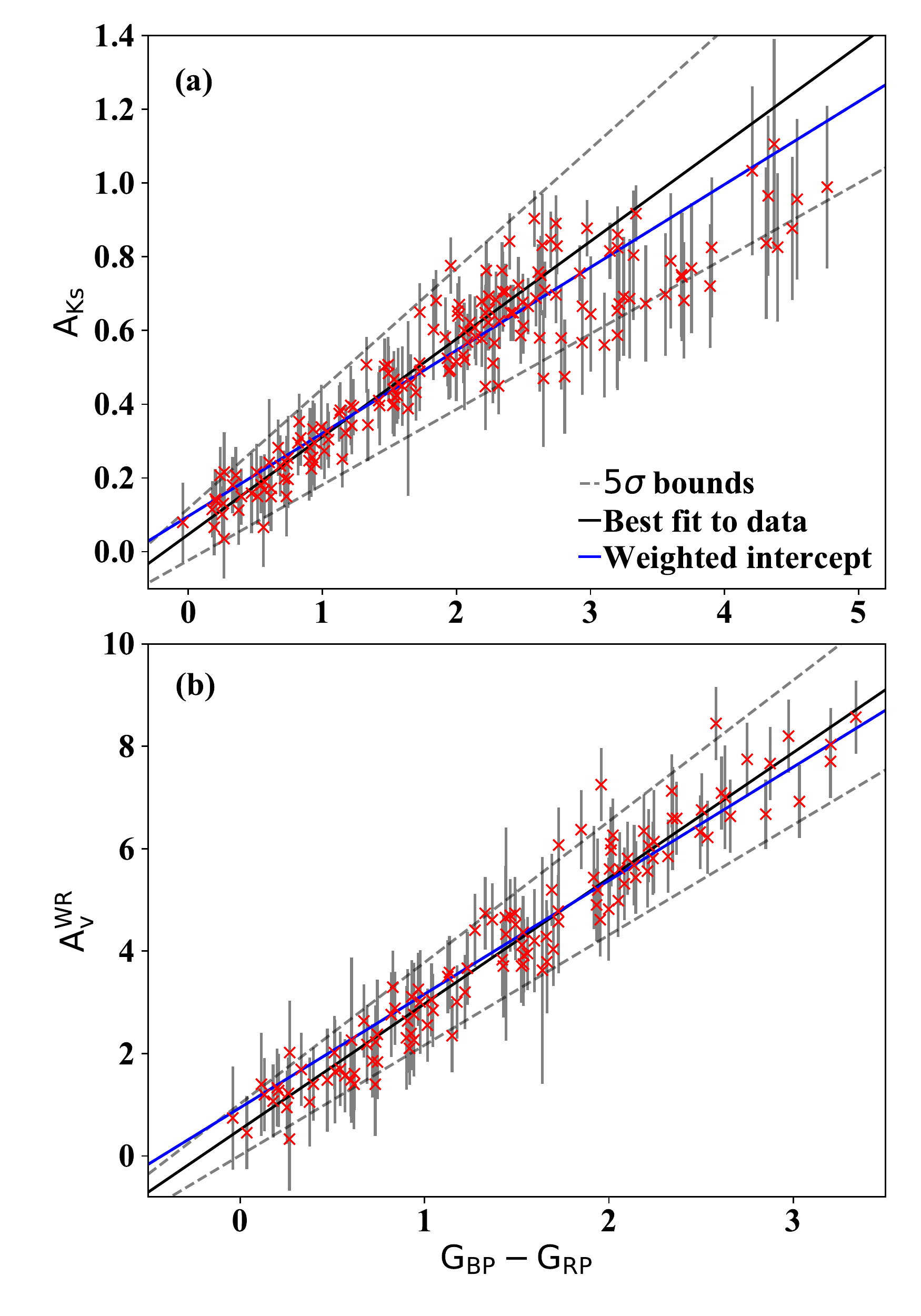}
  \caption{\textit{Gaia} $G_{BP}-G_{RP}$ colours for Galactic WR stars compared to (a) $\mathrm{K_s}$-band and (b) $\mathrm{v^{WR}}$ band extinctions. In (a), the solid black line presents the best fit to data with $G_{BP}-G_{RP}$<3 while in (b), the solid line is a best fit to all data. The grey dashed lines are the 5$\sigma$ bounds, based on the uncertainties of the fit parameters. The solid blue line is also the best fit to the data, but weighted so that it passes through $A^{\rm WR}_{v}=A_{Ks}$=0 at $(G_{BP}-G_{RP})_0=-0.43$, as expected for a generic B0\,V star.} 
	\label{fig:A_bp_rp}
\end{figure}

Since dust extinction preferentially attenuates blue {\textbf wavelengths}, the \textit{Gaia} $G_{BP}-G_{RP}$ can be used as a proxy for extinction. Some stars had unusually high $\mathrm{K_s}$ band extinctions (possibly due to incorrect photometry), which led to erroneous absolute magnitudes. Figure ~\ref{fig:A_bp_rp}(a)
shows the relationship between $(G_{BP}-G_{RP})$ and $A_{Ks}$, while Fig~\ref{fig:A_bp_rp}(b) compares $(G_{BP}-G_{RP})$ and $A\substack{WR \\ v}$. A 5$\sigma$ (grey dashed lines) cut-off from the line of best fit (black solid line) was used to exclude incorrect extinctions. Some values of $A\substack{WR \\ v}$ were also excluded for being outliers, indicating an issue either with some photometry or the $G_{BP}-G_{RP}$ magnitudes.

To obtain meaningful results at low $G_{BP}-G_{RP}$ (where we have no observations) we ensure that the extinction is zero at the intrinsic colour, $(G_{BP}-G_{RP})_0$. We obtain $(G_{BP}-G_{RP})_0$ for a generic blue energy distribution, namely a B0\,V spectral type, with $V-I$=$-$0.44 in the Johnson filter \citep{2001ApJ...558..309D}. We transform this relation to the Cousins system \citep{1979PASP...91..589B} and finally to $(G_{BP}-G_{RP})_0=-0.43$, using the $V-I$ to $G_{BP}-G_{RP}$ calibration in \citet{2018A&A...616A...4E}.

Carrasco \& Jordi (priv. comm) (using methodology from \citealt{2010A&A...523A..48J}) provide the transformation from $A_V$ to $A_G$ by artificially reddening template PoWR WR spectra with different extinctions (from $A_V\sim0.5$ to 36 mag). Synthetic photometry for the \textit{Gaia} \citep{2018A&A...619A.180M} passbands was then obtained at each $A_V$. This allowed for the calculation of $E(G_{BP}-G_{RP})$ and $A_G$. The results from Carrasco \& Jordi allow us to find the intrinsic colour $(G_{BP}-G_{RP})_0$ for each WR subtype. The generic B0\,V model we have used to calculate $(G_{BP}-G_{RP})_0$, is within the uncertainty of the average WR value $(G_{BP}-G_{RP})_0=-0.35\pm0.14$ of the subtypes in Table ~\ref{table:intrinsic_col}. 

\begin{table}
  \caption{Conversion equations between narrowband $\mathrm{v^{WR}}$ and \textit{Gaia} G band filters for $(G_{BP}-G_{RP})_0$ of different spectral types, using results from Carrasco \& Jordi (valid for $A_v$ < 12).}
  \begin{tabular}{ccc}
  \hline
  WR class & $(G_{BP}-G_{RP})_0$ & $A\substack{WR \\ v}$ to $A_G$ \\
  \hline
  WNE-w & $-$0.421 & $-$0.0169$A_v\mathrm{^{2}}$+0.894$A_v$ \\
  WNE-s & $-$0.136 & $-$0.0159$A_v\mathrm{^{2}}$+0.871$A_v$ \\
  WN6ha & $-$0.406 & $-$0.0166$A_v\mathrm{^{2}}$+0.891$A_v$ \\
  WN8 & $-$0.163 & $-$0.0157$A_v\mathrm{^{2}}$+0.868$A_v$ \\
  WN9 & $-$0.359 & $-$0.0163$A_v\mathrm{^{2}}$+0.886$A_v$ \\
  WC5 & $-$0.619 & $-$0.0178$A_v\mathrm{^{2}}$+0.933$A_v$ \\
  WC7 & $-$0.479 & $-$0.0182$A_v\mathrm{^{2}}$+0.921$A_v$ \\
  WC8 & $-$0.360 & $-$0.0178$A_v\mathrm{^{2}}$+0.901$A_v$ \\
  WC9 & $-$0.159 & $-$0.0156$A_v\mathrm{^{2}}$+0.870$A_v$ \\
  B0V SED & $-$0.430 &  \\
  \hline
  \end{tabular}
  \label{table:intrinsic_col}
\end{table} 

\renewcommand{\arraystretch}{1.5}
\begin{table}
  \caption{Average absolute magnitudes for Galactic Wolf-Rayet subtypes in $\mathrm{v^{WR}}$ and $\mathrm{K_s}$ band filters. In the $\mathrm{v^{WR}}$ band, the WC9d sample has been combined with non dusty WC9 stars.}
  \begin{tabular}{|p{.1\textwidth}|p{.08\textwidth}|p{.05\textwidth}|p{.08\textwidth}|p{.05\textwidth}|p{.08\textwidth}|p{.08\textwidth}|p{.1\textwidth}|p{.1\textwidth}|p{.08\textwidth}|p{.05\textwidth}|}
  \hline
  WR subtypes & $M_{v^{WR}}$ (mag) & N($\mathrm{v^{WR}}$) & $M_{Ks}$ (mag) & N($\mathrm{K_s}$) \\
  \hline
  WN3-4 & $-3.6\pm0.5$ & 6 & $-3.1\pm0.6$ & 7 \\
  WN5-6 & $-4.3\pm0.6$ & 22 & $-4.0\pm0.5$ & 33 \\
  WN6-7ha & $-6.5\pm0.3$ & 3 & $-6.2\pm0.3$ & 5 \\
  WN4-6b & $-4.5\pm0.6$ & 13 & $-4.6\pm0.7$ & 15 \\
  WN7 & $-4.6\pm0.6$ & 10 & $-4.8\pm0.3$ & 15 \\
  WN8 & $-5.7\pm0.6$ & 8 & $-6.0\pm0.8$ & 13 \\
  WN8-9ha & $-7.0\pm0.4$ & 2 & $-6.8\pm0.4$ & 2 \\
  WN9 & $-6.0\pm0.8$ & 2 & $-5.7\pm0.7$ & 6 \\
  Of/WN & $-5.8\pm0.1$ & 2 & $-6.1\pm0.1$ & 3 \\
  WO2-4 & $-3.1\pm1.4$ & 3 & $-2.6\pm1.0$ & 4 \\
  WC4-5 & $-4.1\pm0.6$ & 11 & $-4.3\pm0.4$ & 11 \\
  WC6-7 & $-3.9\pm0.4$ & 19 & $-4.9\pm0.4$ & 22 \\
  WC8 & $-4.5\pm0.9$ & 6 & $-5.3\pm0.5$ & 7 \\
  WC9 & $-4.6\pm0.4$ & 12 & $-4.8\pm0.5$ & 9 \\
  WC9d &  &  & $-6.6\pm0.8$ & 13 \\
  \hline
  \end{tabular}
  \label{table:avg_absmag}
\end{table}

For the $\mathrm{K_s}$ band, we obtain the $G_{BP}-G_{RP}$ to $A_{Ks}$ relationship using data with $G_{BP}-G_{RP}<3$. This is the regime in which $A_{Ks}$ follows the extinction law, as these stars are also observed in the $\mathrm{v^{WR}}$ band. At higher $G_{BP}-G_{RP}$, the calculated extinction begins to deviate from this relationship. The empirical fit is shown in blue in Figure ~\ref{fig:A_bp_rp}(a) and has the form:
 \begin{equation} \label{eq:K_vs_bp_rp}
  A = X(G_{BP}-G_{RP})+Y
\end{equation} 
where $G_{BP}-G_{RP}$ is the value from the \textit{Gaia} catalogue, $X$=0.2250 and $Y$=0.0961. The $\mathrm{v^{WR}}$ band, shown in Figure ~\ref{fig:A_bp_rp}(b), was much more closely grouped around the line of best fit, with $X$=2.217 and $Y$=0.9436. The gradient is 9.85 times the gradient for the $\mathrm{K_s}$ band. This is slightly larger than the $A_{Ks} = A\substack{WR \\ v}/9.35$ extinction law used to calculate values of $A_{Ks}$ with $A\substack{WR \\ v}$. The deviation reflects the fact that some values of $A_{Ks}$ were not calculated using that extinction law. 

\begin{figure*}
	\begin{adjustwidth}{-2.7cm}{2.7cm}
  \includegraphics[width=1.3\linewidth]{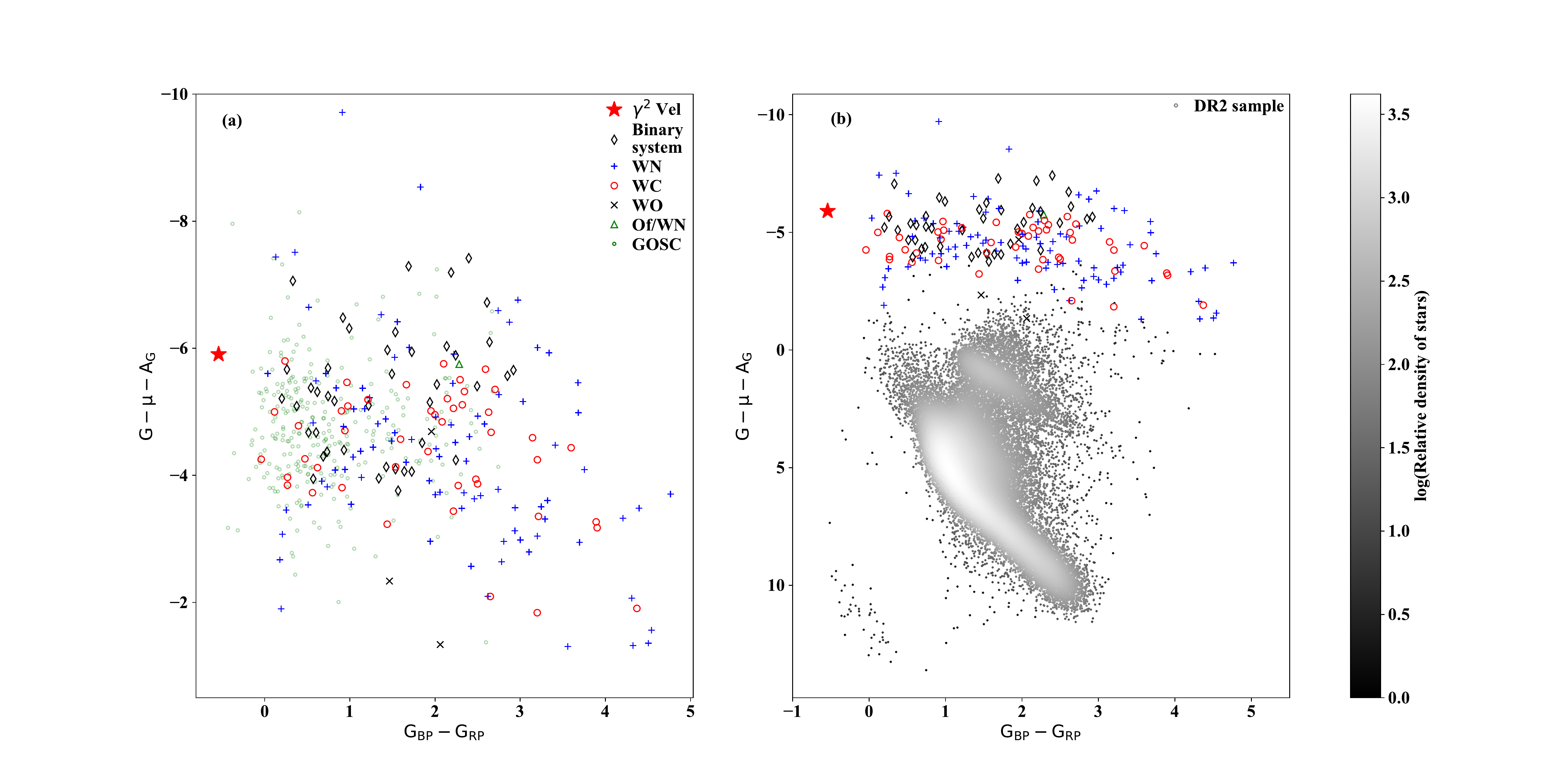}
  \end{adjustwidth}
        \caption{(a) \textit{Gaia} DR2 colour magnitude diagram for Galactic WR stars plus O stars from GOSC (v4.1, \citealt{2013msao.confE.198M}). Absolute magnitudes are calculated using our inferred distance moduli $\mu$ and $A_G$ (converted from $A^{\rm WR}_{v}$ using the relation from Carrasco \& Jordi). The red star is the WR component of $\gamma$ Velorum, the only WR star with a trigonometric parallax from \textit{Hipparcos}; (b) \textit{Gaia} DR2 colour magnitude diagram for Galactic WR stars plus 70,000 stars from DR2, satisfying the selection criteria from section 2.1 of \citet{2018A&A...616A..10G}.}
	\label{fig:cmd}
\end{figure*}

We can also use the synthetic photometry from Carrasco \& Jordi to calculate the conversion relationship from $A\substack{WR \\ v}$ to $A_G$ (also shown in Table ~\ref{table:intrinsic_col}), by converting $A_V$ in their relationship to $A\substack{WR \\ v}$. This enables us to calculate the absolute \textit{Gaia} G magnitude and present the \textit{Gaia} colour magnitude diagram (CMD) in Figure ~\ref{fig:cmd}, for the most reliable WR results. Fig.~\ref{fig:cmd}(a) presents a CMD for Galactic WR stars plus visually bright O stars from v4.1 of the Galactic O Star Catalogue (GOSC, \citealt{2013msao.confE.198M}), while Fig.~\ref{fig:cmd}(b) compares the CMD of WR stars to 70,000 DR2 stars from \citet{2018A&A...616A..10G}. Two exceptionally bright stars are the extreme hypergiants He\,3-519 (WR31a) and AG Car (WR31b), which exhibit very late WN characteristics at extreme visual minima \citep{1994A&A...281..833S}.

\subsection{Absolute magnitudes} \label{ssec:absmag}

\begin{figure*}
  \begin{minipage}[c]{0.8\linewidth}
      \includegraphics[width=\linewidth]{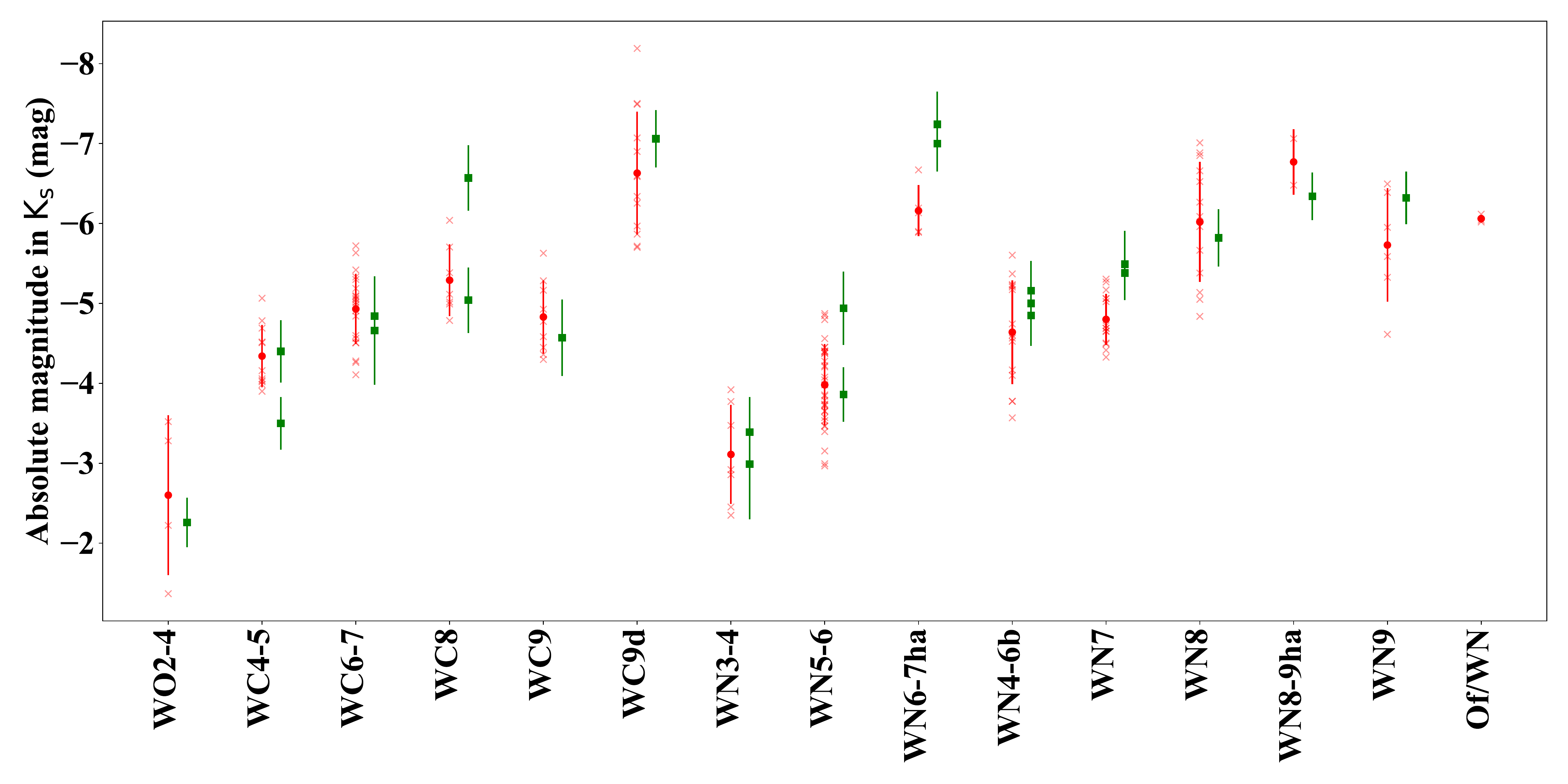}
  \end{minipage}\hfill
  \begin{minipage}[c]{0.2\linewidth}
      \caption{Absolute magnitudes in the $\mathrm{K_s}$ band. Red crosses are individual WR star results and the red circle is the average for each spectral subtype (with the sample standard deviation of the data as the uncertainties). Green squares are the comparative data from \citet{2015MNRAS.447.2322R}.} 
      \label{fig:MK}
  \end{minipage}
\end{figure*}

\begin{figure*}
  \begin{minipage}[c]{0.8\linewidth}
      \includegraphics[width=\linewidth]{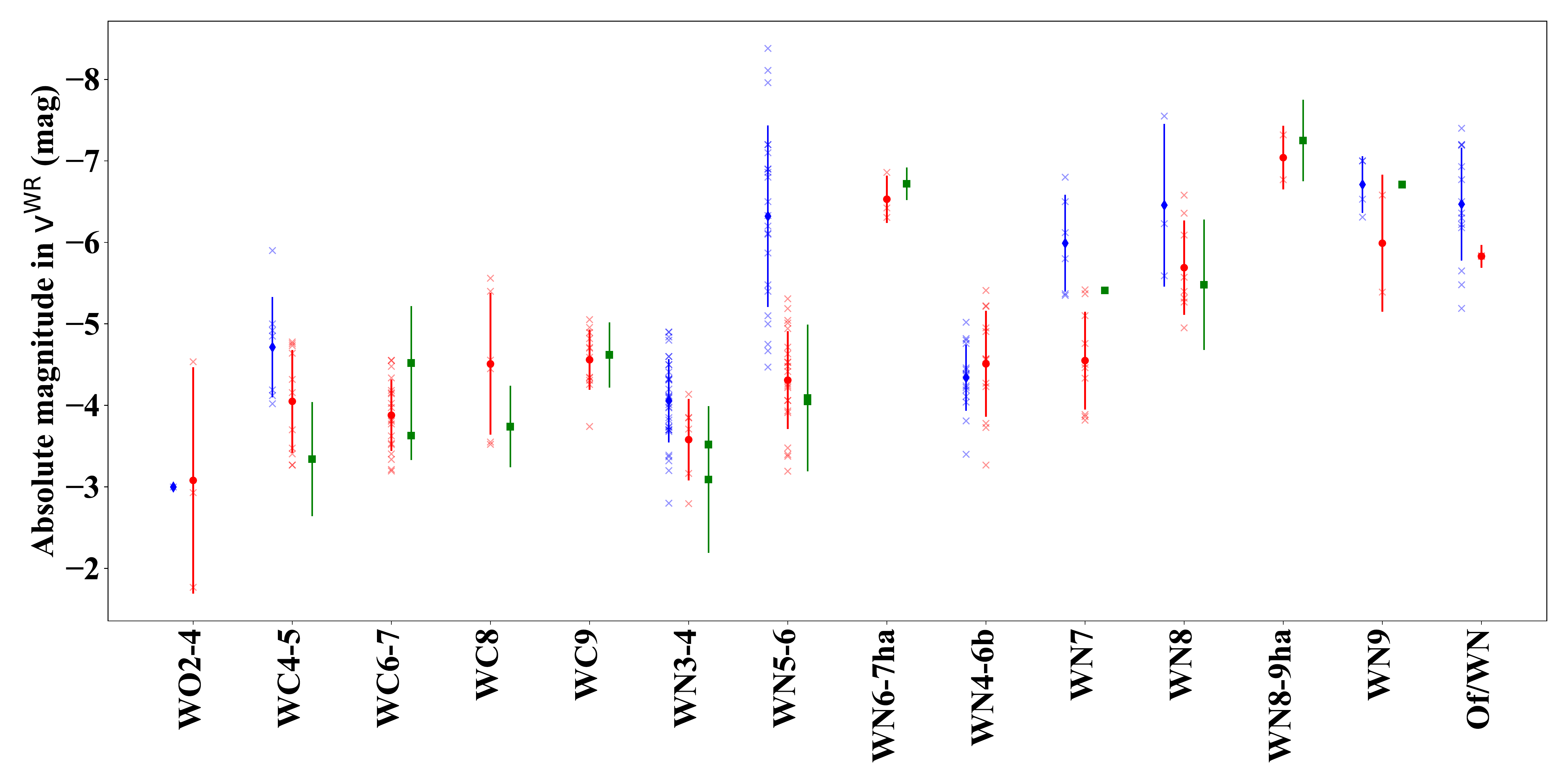}
  \end{minipage}\hfill
  \begin{minipage}[c]{0.2\linewidth}
      \caption{Absolute magnitudes in the $\mathrm{v^{WR}}$ band. Red crosses are individual WR star results and the red circle is the average for each spectral subtype (with the sample standard deviation of the data as the uncertainties). Green squares are the comparative data from \citet{2001NewAR..45..135V}. Results from the LMC (\citealt{2014A&A...565A..27H}, \citealt{2019A&A...627A.151S} and \citealt{2002A&A...392..653C}) are shown in blue, with crosses for individual stars and the diamond the average for each subtype. LMC WN5-6 stars include very luminous H-rich main sequence WN5--6h stars. Results for WO were calculated using \citet{2015A&A...581A.110T} and \citet{1988AJ.....96.1076T}} 
      \label{fig:Mv}
  \end{minipage}
\end{figure*}

We used the extinctions, distances and apparent magnitudes to calculate the absolute magnitudes for stars that have reliable extinctions (within the 5$\sigma$ bounds of Figure ~\ref{fig:A_bp_rp}). Repeating the calculation using a Monte Carlo selection (bootstrapping with replacement) from the distributions of the three parameters, produced a binned histogram of absolute magnitude against frequency. This acted as a proxy for the probability distribution of each absolute magnitude. A Gaussian or Weibull distribution was fit to the binned data, to find the most likely absolute magnitude and uncertainties (more details are available in Appendix F of the online material).

For binaries, the absolute magnitudes of Wolf-Rayet components were separated from the total system magnitude. 

A multi step process of sigma clipping allowed us to find reliable absolute magnitudes for all WR stars. First, stars with high astrometric excess noise, or unrealistically low absolute magnitudes ($\geq -$1 mag) were removed from the sample. We then calculated the averages of the remaining stars in each subtype class. Stars with unusually high or low absolute magnitudes (defined as were greater than one sample standard deviation, from the mean) were then cut from the sample. This cut-off provided a good balance between excluding clearly incorrect values and including valid ones across all subtypes. 

The remaining sample contained only the most reliable absolute magnitude results in each subclass and were used to calculate the averages presented in Table ~\ref{table:avg_absmag}. 
LBVs, aside from He~3-519 (WR31a) and AG Car (WR31b), were removed due to variability. WR20-2, WR42-1, WR43-2, WR43-3 were also excluded from the averages, owing to uncertain subtypes.

We obtain $\mathrm{K_s}$ band results for dusty subtypes (WC8d and WC9d) by converting $A\substack{WR \\ v}$ to $A_{Ks}$, using the standard extinction law. This method prevents the IR dust emission from contaminating the extinction calculation. The absolute magnitudes could then be calculated for each subtype and in each filter, with the standard deviation providing upper and lower bounds on the typical absolute magnitudes. The WC9d were combined in the $\mathrm{v^{WR}}$ band, but not in the $\mathrm{K_s}$ band, as their IR excess renders them brighter than dust free WR stars. As there were only three WC8d (WR48a, WR53 and WR113) in the final sample, these stars were grouped with the non dusty WC8 stars and only WR113 was used to calculate the final absolute $\mathrm{K_s}$ in Table~\ref{table:avg_absmag}. Excluding WR113 from the average, we obtain $M_{Ks}$=--5.3~mag for WC8 stars, the same result as Table~\ref{table:avg_absmag}.

\begin{table*} 
  \renewcommand{\arraystretch}{1.5}
  \caption{Absolute $\mathrm{K_s}$-band magnitudes for Galactic WR stars. The full table is available in the online supplementary material.}
  \begin{tabular}{|p{.07\textwidth}|p{.12\textwidth}|p{.05\textwidth}|p{.05\textwidth}|p{.05\textwidth}|p{.05\textwidth}|p{.08\textwidth}|p{.05\textwidth}|p{.08\textwidth}|p{.08\textwidth}|p{.05\textwidth}|}
  \hline
  WR Number & Spectral type & $\mathrm{K_s}$ (mag) & $\mu$ (mag) & J$-\mathrm{K_s}$ (mag) & H$-\mathrm{K_s}$ (mag) & $A_{Ks}$ (mag) & $M\substack{\mathrm{Sys} \\ \mathrm{\mathrm{K_s}}}$ (mag) & F$\substack{\mathrm{WR} \\ \mathrm{K_s}}$/ F$\substack{\mathrm{Sys} \\ \mathrm{K_s}}$ & $M\substack{\mathrm{WR} \\ \mathrm{K_s}}$ (mag) & Flags \\
  \hline
  WR1 & WN4b & 7.48 & 12.49 & 0.73 & $0.38$ & 0.30$\pm$0.08 &  &  & $-5.4\substack{+0.3 \\ -0.3}$ & b: \\
  WR3 & WN3ha & 10.01 & 12.31 & 0.23 & $0.12$ & 0.11$\pm$0.08 &  &  & $-2.5\substack{+0.3 \\ -0.4}$ & b: \\
  WR4 & WC5+? & 7.88 & 12.87 & 0.87 & $0.69$ &  0.18$\pm$0.11 &  &  & $-5.2\substack{+0.4 \\ -0.5}$ & b \\
  WR5 & WC6 & 7.65 & 12.36 & 0.98 & $0.69$ & 0.30$\pm$0.11 &  &  & $-5.1\substack{+0.3 \\ -0.3}$ & g \\
  WR6 & WN4b & 5.89 & 11.78 & 0.46 & $0.34$ & 0.05$\pm$0.08 &  &  & $-6.0\substack{+0.3 \\ -0.4}$ & b \\
  WR7 & WN4b & 9.27 & 13.13 & 0.70 & $0.40$ & 0.24$\pm$0.08 &  &  & $-4.2\substack{+0.4 \\ -0.5}$ & g \\
  WR8 & WN7o/CE & 7.93 & 12.87 & 0.64 & $0.39$ & 0.31$\pm$0.08 &  &  & $-5.3\substack{+0.3 \\ -0.3}$ & b: \\
  WR9 & WC5+O7 & 7.54 & 13.3 & 0.91 & $0.57$ & 0.45$\pm$0.08 & $-6.3\substack{+0.3 \\ -0.4}$ & $0.60\pm0.24$ & $-5.7\substack{+0.9 \\ -0.7}$ & b \\
  WR10 & WN5h & 9.61 & 13.69 & 0.44 & $0.28$ & 0.24$\pm$0.17 &  &  & $-4.4\substack{+0.5 \\ -0.5}$ & g \\
  \hline
  \label{table:final_absmagk}
   \end{tabular}
Columns: (1) WR Number, (2) Spectral type, (3) $\mathrm{K_s}$ apparent magnitude, (4) Distance modulus $\mu$, (5) J$-\mathrm{K_s}$ colour, (6) H$-\mathrm{K_s}$ colour, (7) $\mathrm{K_s}$ band extinction $A_{Ks}$, (8) Absolute magnitude of binary system (including companion), (9) Fraction of light contributed to the binary system by the WR component, (10) Absolute magnitude of WR star, (11) Error flags, where $M$ > upper$_{\rm initial}$ or $M$ < lower$_{\rm initial}$ = b, $M$ > upper$_{\rm final}$ or $M$ < lower$_{\rm final}$ = b: ($_{\rm initial}$ denotes the averages calculated before sigma clipping, $_{\rm final}$ are the final absolute magnitude boundaries) and g are results with no issues.
\end{table*}

\begin{table*} 
  \renewcommand{\arraystretch}{1.5}
  \caption{Absolute $\mathrm{v^{WR}}$-band magnitudes for Galactic WR stars. The full table is available in the online supplementary material.}
  \begin{tabular}{|p{.07\linewidth}|p{.12\linewidth}|p{.06\linewidth}|p{.05\linewidth}|p{.07\linewidth}|p{.08\linewidth}|p{.08\linewidth}|p{.12\linewidth}|p{.05\linewidth}|p{.05\linewidth}|}
  \hline
  WR Number & Spectral type & $\mathrm{v^{WR}}\pm$0.1 (mag) & $\mu$ (mag) & $\mathrm{(b-v)}^{WR}$ (mag) & $A\substack{WR \\ v}$ (mag) & $M\substack{\mathrm{Sys} \\ \mathrm{v}}$ (mag) & F$\substack{\mathrm{WR} \\ \mathrm{v}}$/ F$\substack{\mathrm{Sys} \\ \mathrm{v}}$ & M$\substack{\mathrm{WR} \\ \mathrm{v}}$ (mag) & Flags \\
  \hline
  WR1 & WN4b & 10.51 & 12.49 & 0.51 & 2.84$\pm$0.71 &  &  & $-4.9\substack{+0.8 \\ -0.8}$ & g \\
  WR3 & WN3ha & 10.70 & 12.31 & -0.06 & 1.07$\pm$0.71 &  &  & $-2.8\substack{+0.8 \\ -0.8}$ & b: \\
  WR4 & WC5+? & 10.53 & 12.87 & 0.20 & 1.65$\pm$1.01 &  &  & $-4.2\substack{+1.1 \\ -1.1}$ & g \\
  WR5 & WC6 & 11.02 & 12.36 & 0.47 & 2.76$\pm$1.01 &  &  & $-4.2\substack{+1.1 \\ -1.1}$ & g \\
  WR6 & WN4b & 6.94 & 11.78 & -0.07 & 0.45$\pm$0.71 &  &  & $-5.4\substack{+0.8 \\ -0.8}$ & b: \\
  WR7 & WN4b & 11.75 & 13.13 & 0.36 & 2.22$\pm$0.71 &  &  & $-3.8\substack{+0.9 \\ -0.9}$ & b: \\
  WR8 & WN7o/CE & 10.48 & 12.87 & 0.47 & 2.88$\pm$0.71 &  &  & $-5.4\substack{+0.8 \\ -0.8}$ & b: \\
  WR9 & WC5+O7 & 10.93 & 13.3 & 0.74 & 4.16$\pm$0.72 & $-6.6\substack{+0.8 \\ -0.8}$ & $0.29\pm0.12$ & $-5.3\substack{+1.4 \\ -1.2}$ & b \\
  WR10 & WN5h & 11.08 & 13.69 & 0.22 & 2.26$\pm$1.61 &  &  & $-5.0\substack{+1.7 \\ -1.7}$ & b: \\
  \hline
  \end{tabular}
  \label{table:final_absmagv}
Columns: (1) WR Number, (2) Spectral type, (3) $\mathrm{v^{WR}}$ apparent magnitude and error, (4) Distance modulus $\mu$, (5) $\mathrm{(b-v)}^{WR}$ colour, (6) $\mathrm{v^{WR}}$ band extinction $A_v$, (7) Absolute magnitude of binary system (including companion), (8) Fraction of light contributed to the binary system by the WR component, (9) Absolute magnitude of WR star, (10) Error flags, where $M$ > upper$_{\rm initial}$ or $M$ < lower$_{\rm initial}$ = b, $M$ > upper$_{\rm final}$ or $M$ < lower$_{\rm final}$ = b: ($_{\rm initial}$ denotes the averages calculated before sigma clipping, $_{\rm final}$ are the final absolute magnitude boundaries) and g are results with no issues.
\end{table*}

In Figure ~\ref{fig:MK}, we present the final absolute $\mathrm{K_s}$ band magnitudes and uncertainties for each subtype. These are compared with corresponding values from \citet{2015MNRAS.447.2322R}. Figure ~\ref{fig:Mv} shows the same distribution for the $\mathrm{v^{WR}}$ band, compared with \citet{2001NewAR..45..135V}. Tables ~\ref{table:final_absmagk} and ~\ref{table:final_absmagv} show results for individual stars (the full lists are in the supplementary online material).

We additionally plot the absolute magnitudes for 116 LMC stars in Figure~\ref{fig:Mv}, using results from \citet{2014A&A...565A..27H} for single WN and Of supergiant stars (excluding WN2b), \citet{2019A&A...627A.151S} for stars in binaries, \citet{2002A&A...392..653C} for single WC stars and reddenings from \citet{2015A&A...581A.110T} and $\mathrm{v^{WR}}$ band magnitudes from \citet{1988AJ.....96.1076T} for BAT99-123 (WO4). We adopt spectral types of LMC late WN stars from \citet{1997A&A...320..500C} instead of \citet{2008MNRAS.389..806S}. 

From Fig.~\ref{fig:Mv}, absolute $\mathrm{v^{WR}}$ magnitudes of LMC stars are brighter than their Galactic analogues, so it is inappropriate to apply LMC WR absolute magnitudes to Galactic stars. LMC WN5--6 stars are particularly bright, since this sample includes the luminous H-rich main sequence WN5--6h stars whose closest Galactic analogues are WN6--7ha stars which are amongst the visually brightest WR stars in the Milky Way.

In total, reasonable absolute magnitudes, extinctions and no \textit{Gaia} excess noise flags, were obtained in 187 cases. 
Absolute magnitudes for almost all WR subtypes revealed standard deviations that overlapped with the uncertainty range of the previous results in both the $\mathrm{v^{WR}}$ and the $\mathrm{K_s}$ bands. The differences between values can be attributed to the improved distance estimates and the increased number of stars with distances. Some stars, such as WR2 (the only WN2 star, \citealt{2019MNRAS.484.5834C}), were not present in the \textit{Gaia} catalogue. 

There is a clear trend across both filters of increasing absolute magnitudes with increasing subtype. In both filters, WN4-6b are brighter than their weak-lined counterparts despite their higher effective temperatures \citep{2006A&A...457.1015H}. WNLha stars are known to be highly luminous, and conform to this expectation. 

\begin{table*}
  \caption{WR stars within 2 kpc of the Sun, including colour excess, K-band extinction and 
    $\mathrm{A_{Ks}}$/kpc, extinction per kpc.}
  \begin{tabular}{|p{.08\linewidth}|p{.15\linewidth}|p{.15\linewidth}|p{.08\linewidth}|p{.05\linewidth}|p{.08\linewidth}|p{.08\linewidth}|p{.08\linewidth}|}
  \hline
  WR Number & Alias & Spectral type & Distance (kpc) & Flags & E(B-V) & $\mathrm{A_{Ks}}$ & $\mathrm{A_{Ks}}$/kpc \\
  \hline
  WR11 & $\gamma$ Vel & WC8+O7.5III-V & 0.34$\substack{+0.04 \\ -0.03}$ & ... & 0.00$\pm$0.30 & 0.00$\pm$0.11 & $0.00\substack{+0.32 \\ -0.32}$ \\
  WR94 & HD 158860 & WN5o & 0.95$\substack{+0.06 \\ -0.06}$ & g & 1.24$\pm$0.21 & 0.45$\pm$0.08 & $0.47\substack{+0.09 \\ -0.08}$ \\
  WR90 & HD 156385 & WC7 & 1.15$\substack{+0.11 \\ -0.09}$ & g & 0.10$\pm$0.30 & 0.04$\pm$0.11 & $0.03\substack{+0.09 \\ -0.03}$ \\
  WR78 & HD 151932 & WN7h & 1.25$\substack{+0.15 \\ -0.12}$ & g & 0.44$\pm$0.21 & 0.16$\pm$0.08 & $0.13\substack{+0.06 \\ -0.06}$ \\
  WR139 & HD 193576 & WN5o+O6III-V & 1.31$\substack{+0.07 \\ -0.06}$ & g & 0.81$\pm$0.24 & 0.30$\pm$0.09 & $0.23\substack{+0.07 \\ -0.07}$ \\
  WR79 & HR 6265 & WC7+O5-8 & 1.37$\substack{+0.12 \\ -0.10}$ & g & 0.31$\pm$0.26 & 0.11$\pm$0.09 & $0.08\substack{+0.07 \\ -0.07}$ \\
  WR145 & AS 422 & WN7o/CE+? & 1.46$\substack{+0.12 \\ -0.10}$ & g & 2.28$\pm$0.39 & 0.83$\pm$0.14 & $0.57\substack{+0.11 \\ -0.10}$ \\
  WR110 & HD 165688 & WN5-6b & 1.58$\substack{+0.15 \\ -0.12}$ & g & 1.13$\pm$0.21 & 0.41$\pm$0.08 & $0.26\substack{+0.05 \\ -0.05}$ \\
  WR111 & HD 165763 & WC5 & 1.63$\substack{+0.32 \\ -0.23}$ & g & 0.22$\pm$0.30 & 0.08$\pm$0.11 & $0.05\substack{+0.07 \\ -0.05}$ \\
  WR142 & Sand 5 & WO2 & 1.65$\substack{+0.11 \\ -0.09}$ & g & 2.13$\pm$0.21 & 0.78$\pm$0.08 & $0.47\substack{+0.06 \\ -0.05}$ \\
  WR105 & NS 4 & WN9h & 1.73$\substack{+0.32 \\ -0.23}$ & g & 2.41$\pm$0.21 & 0.88$\pm$0.08 & $0.51\substack{+0.10 \\ -0.08}$ \\
  WR134 & HD 191765 & WN6b & 1.75$\substack{+0.13 \\ -0.11}$ & g & 0.46$\pm$0.21 & 0.17$\pm$0.08 & $0.10\substack{+0.04 \\ -0.04}$ \\
  WR52 & HD 115473 & WC4 & 1.75$\substack{+0.16 \\ -0.13}$ & g & 0.59$\pm$0.30 & 0.22$\pm$0.11 & $0.12\substack{+0.06 \\ -0.06}$ \\
  WR144 & HM19-1 & WC4 & 1.75$\substack{+0.24 \\ -0.19}$ & g &  & 0.47$\pm$0.19 & $0.27\substack{+0.11 \\ -0.11}$ \\
  WR93 & Th10-19 & WC7+O7-9 & 1.76$\substack{+0.19 \\ -0.15}$ & g & 1.67$\pm$0.23 & 0.61$\pm$0.08 & $0.34\substack{+0.06 \\ -0.06}$ \\
  WR142-1 & HBHalpha 4203-27 & WN6o & 1.77$\substack{+0.23 \\ -0.18}$ & g &  & 0.69$\pm$0.16 & $0.39\substack{+0.10 \\ -0.10}$ \\
  WR113 & HD 168206 & WC8d+O8-9IV & 1.80$\substack{+0.24 \\ -0.19}$ & g & 0.94$\pm$0.21 & 0.34$\pm$0.08 & $0.19\substack{+0.05 \\ -0.05}$ \\
  WR142a & PCG02 1 & WC8 & 1.81$\substack{+0.61 \\ -0.37}$ & g &  & 0.83$\pm$0.19 & $0.46\substack{+0.19 \\ -0.14}$ \\
  WR133 & HD 190918 & WN5o+O9I & 1.85$\substack{+0.16 \\ -0.14}$ & g & 0.36$\pm$0.21 & 0.13$\pm$0.07 & $0.07\substack{+0.04 \\ -0.04}$ \\
  WR113-2 & SMG09 1425$\_$47 & WC5-6 & 1.86$\substack{+0.90 \\ -0.56}$ & g &  & 0.65$\pm$0.21 & $0.35\substack{+0.21 \\ -0.16}$ \\
  WR70-5 & WM10 11b & WC9 & 1.95$\substack{+0.75 \\ -0.47}$ & g &  & 1.26$\pm$0.26 & $0.65\substack{+0.28 \\ -0.21}$ \\
  WR98 & HDE 318016 & WN8o/C7 & 1.96$\substack{+0.31 \\ -0.24}$ & g & 1.59$\pm$0.21 & 0.58$\pm$0.08 & $0.29\substack{+0.06 \\ -0.05}$ \\
  WR25 & HD 93162 & O2.5If*/WN6+O & 1.97$\substack{+0.18 \\ -0.15}$ & g & 0.93$\pm$0.32 & 0.34$\pm$0.11 & $0.17\substack{+0.06 \\ -0.06}$ \\
  WR135 & HD 192103 & WC8 & 1.98$\substack{+0.18 \\ -0.15}$ & g & 0.41$\pm$0.21 & 0.15$\pm$0.08 & $0.08\substack{+0.04 \\ -0.04}$ \\
  WR85 & HD 155603B & WN6h & 1.99$\substack{+0.30 \\ -0.24}$ & g & 1.03$\pm$0.21 & 0.37$\pm$0.08 & $0.19\substack{+0.05 \\ -0.04}$ \\
  \hline
  \end{tabular}
  \label{table:in_2kpc}
\end{table*}

The spread in absolute magnitudes is similar to those previously obtained in the near-IR, but slightly larger in the $\mathrm{v^{WR}}$ band. \citet{2015MNRAS.447.2322R} quote a range of 0.3-0.6 mag, whilst the standard deviation in our $\mathrm{K_s}$ band results spans 0.1-1.0 mag, but is also more typically 0.3-0.6 mag. For the $\mathrm{v^{WR}}$ band, the standard deviations range from 0.3-1.4 mag and mostly have standard deviations between 0.4-0.6 mag.

We therefore corroborate the  findings of \citet{2019A&A...621A..92S} that WC stars of the same subtype have a broader range of absolute magnitudes than expected. We also posit this is true for WN stars (\citealt{2019A&A...625A..57H} also note the relations between absolute magnitude and subtype are not strict). The uncertainties show no systematic differences between WC and WN classes or regular variation across subtypes. However, particularly in the $\mathrm{v^{WR}}$ band, some classes suffered from very small numbers of WR stars (only 2 WN9 stars had $\mathrm{v^{WR}}$ band magnitudes, for instance). This increases the size of the uncertainties on the mean result.

Due to this intrinsic variation, we advise caution when using averages as absolute magnitude calibrations and recommend accounting for the large uncertainties by exploring other methods, such as a Bayesian approach with a probability distribution centred on the average magnitude. We also recommend continued use of the intrinsic colours in Table ~\ref{table:powr_types}, rather than calculating new values using our methods and results. The large uncertainties of our absolute magnitudes, mean that propagated uncertainties of any resulting intrinsic colours are correspondingly large. These new uncertainties are far larger than the intrinsic colours from Table ~\ref{table:powr_types}.

\subsection{Sensitivity of results to adopted intrinsic colours}

We test the sensitivity of the results to the intrinsic colours. For the $v^{WR}$ band, this is straightforward in that any difference in ${\mathrm{(b-v)_0^{WR}}}$ is propagated through to the extinction (so multiplied by 4.12, \citealt{1982IAUS...99...57T}). However, within the $\mathrm{K_s}$ band, the combination of (J$-\mathrm{K_s}$)$_0$ and (H$-\mathrm{K_s}$)$_0$ complicates this somewhat and we test the effects by calculating $M_{Ks}$ with alternative J$-\mathrm{K_s}$ and H$-\mathrm{K_s}$ synthetic colours. These are taken from the PoWR grids (\citealt{2004A&A...427..697H} and \citealt{2015A&A...579A..75T} for WN, \citealt{2012A&A...540A.144S} for WC), using the same models as Table ~\ref{table:powr_types}. Unlike the $\mathrm{b-v}^{WR}$ colours, these are only valid at the monochromatic wavelengths and not the whole filter bands, which are affected by emission lines, especially for early-type WC stars. The difference in absolute magnitudes are between 0.05 for WN5-6 and 0.2 for WC6-7 and WN2-4 (as emission lines fall within the filter band and are not included in the monochromatic result), with most subtypes falling between 0.1 and 0.2. In all instances, this was well within the uncertainties on individual magnitudes. 

\subsection{Photometric Flags} \label{ssec:pflag}

In addition to the \textit{Gaia} flag, we identify results with potentially spurious absolute magnitudes. As stars with incorrect extinctions were removed, spurious results can indicate either incorrect apparent magnitudes, or an incorrect \textit{Gaia} parallax, whose distance generates the wrong absolute magnitude. We therefore adopt two different flags, one where the absolute magnitude is implausible and another where the absolute magnitude only just falls outside the uncertainty of the subtype average. The latter does not necessarily indicate a bad result, but these data should be treated with caution.  

M > upper$_{initial}$ or M < lower$_{initial}$ = b

M > upper$_{final}$ or M < lower$_{final}$ = b:

where upper and lower are the upper and lower magnitude bounds of the absolute magnitude average. $_{initial}$ denotes the averages calculated before sigma clipping (Section ~\ref{sec:absmag}), $_{final}$ are the final absolute magnitude boundaries (as in Table ~\ref{table:avg_absmag}) and M is the absolute magnitude of individual WR stars. Results with a 'b' flag are highly implausible and lie well outside the range of acceptable absolute magnitudes, whilst those with a 'b:' flag are still acceptable, but fall outside the 1$\sigma$ uncertainties of the results in Table ~\ref{table:avg_absmag}. Again, results without any of these issues are given the 'g' flag. Results without any absolute magnitudes are flagged with 'u'. These stars were included to provide the reader with the distance moduli of the stars and any other helpful information (e.g apparent magnitudes), if their absolute magnitudes could not be calculated.

For all subsequent analysis we use only the most photometrically reliable results, which have a 'b:' or 'g' flag in either the $\mathrm{v^{WR}}$ band, or the $\mathrm{K_s}$ band. These data do not have high astrometric excess noise ('a') Gaia data quality flags. Results with, for example, two 'b' flags were excluded. These flags are applied to the absolute magnitudes in Tables ~\ref{table:final_absmagv} and ~\ref{table:final_absmagk}.

We note that 13 objects retained in this selection process had either negative parallax ('n') or high parallax to error ratio ('e') \textit{Gaia} flags. However, the reliable absolute magnitudes mean the distances may still be valid. 


\section{New distances to WR stars and comparison to other \textit{Gaia} derived distances} \label{sec:distdisc}

We can compare the WR star sample from \textit{Gaia} to the total population. There is no substantial difference between the latitude and longitude distribution of WR stars detected in \textit{Gaia} and the total known WR distribution. The exception is for some regions, such as around Westerlund 1 and towards the Galactic Centre, which went undetected by \textit{Gaia} due to their high extinctions (with $A_V>30$ mag in the latter case). 

Crowding presented an additional challenge. WR 43A and 43B are not included in the final distance catalogue as the same \textit{Gaia} source was detected for both stars. The detection for WR43C is also spurious, as the position overlaps with other objects. These stars are located in the compact cluster NGC~3603 (\citealt{2008AJ....135..878M}, \citealt{1998MNRAS.296..622C}) and therefore blending is to be expected. It is possible that further stars are missing parallaxes due to crowding, as this issue would reduce the quality of the \textit{Gaia} five parameter solution below acceptable limits, and cause it to be excluded from the \textit{Gaia} catalogue.

Finally, some stars may not have been detected due to their close binary companions. \citet{2018A&A...616A..17A} shows that completeness falls for separations below 2'', to a limit at 0.12''.  This may account for three missing stars with narrowband $\mathrm{v^{WR}}$ < 15 mag (WR2, WR63 and WR86), two of which (WR63 and WR86) have known companions. 

Table ~\ref{table:final} includes distances for each WR star with measured parallaxes. Also included are the 68\% credible intervals. Table ~\ref{table:in_2kpc} lists the closest WR stars (with reliable results) within 2 kpc of the Sun. We find 25 WR stars within this distance, similar to the 30 WR stars within 2 kpc from \citet{1983ApJ...274..302C}. We also calculate distances to O stars using our Bayesian prior and GOSC v4.1 \citep{2013msao.confE.198M}. For the O star population within 2 kpc, we obtain a WR/O ratio of 0.09. This ratio is within the 0.07--0.10 range of \citet{1983ApJ...274..302C}, found by comparing lifetimes of H and He core burning phases from massive star models, as an analogue to O star and WR star phases. However, our ratio includes all O stars, and not just the most massive population that WR stars are descended from. \citet{1983ApJ...274..302C} also calculate a WR/O ratio with only O stars >40\Msolar, and find a much higher ratio of 0.36$\pm$0.15.

Table~\ref{table:in_2kpc} also includes $K_s$-band extinctions, and extinctions per kpc for these nearby WR stars, with ${A_{K_s}}$/kpc $\sim$ 0.26 mag, albeit with significant star-to-star variation. Dust extinctions of stars in common with the 3D dust map from Pan-STARRS1 and 2MASS \citet{2015ApJ...810...25G} shows reasonable overall agreement.

\begin{figure}
	\centering
	\includegraphics[width=\linewidth]{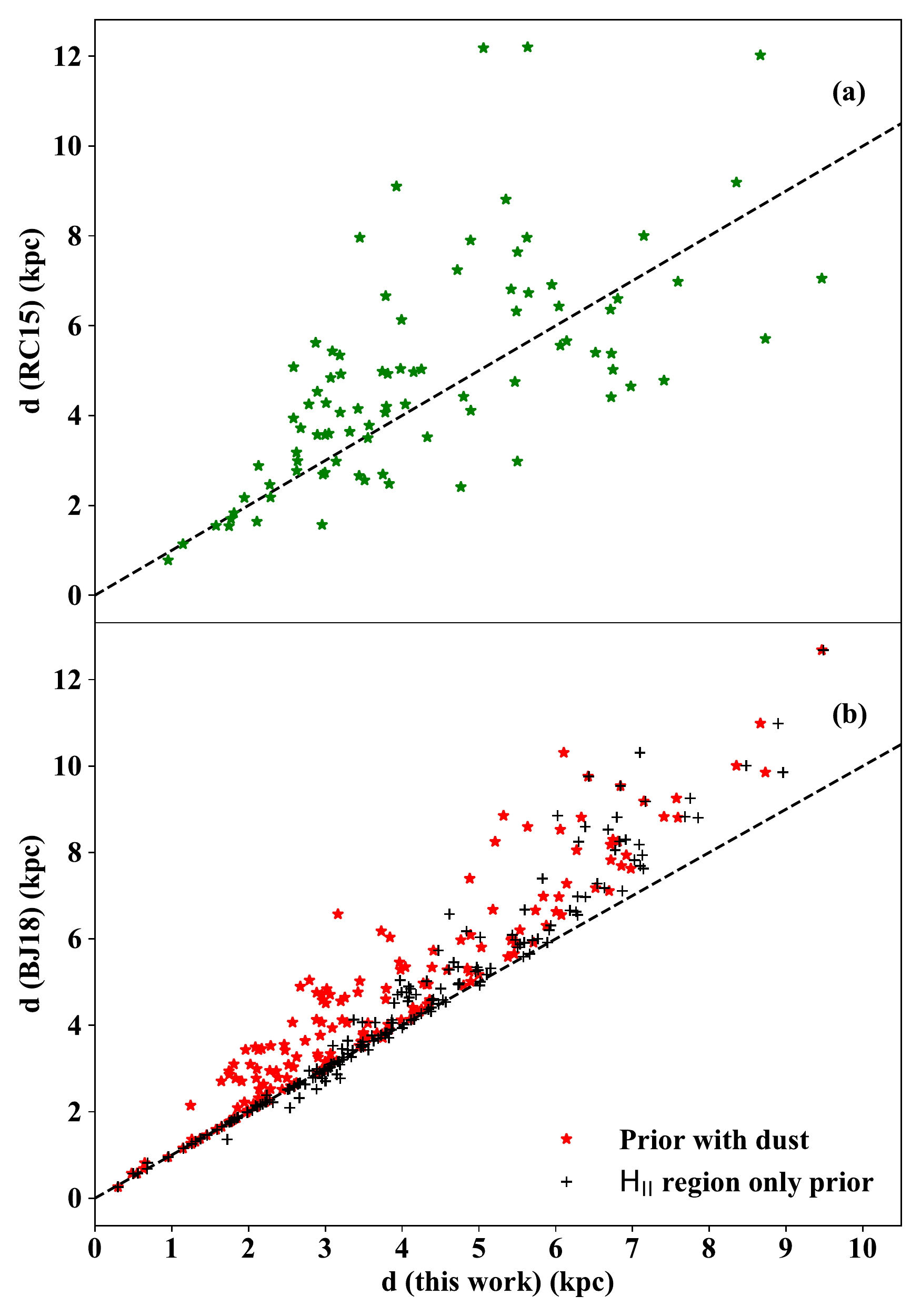}
    \caption{(a) A comparison between distances to Galactic WR stars in common between this work and \citet{2015MNRAS.447.2322R}. The black dashed line indicates one-to-one agreement. Error bars from \citet{2015MNRAS.447.2322R} have been omitted for clarity; (b) A comparison between WR distances obtained in this work and \citet{2018AJ....156...58B}. We illustrate the effect of extinction by presenting the full prior including both dust and \hii regions (red stars) and a prior with only \hii regions (black cross).} 
	\label{fig:cf_ori}
\end{figure}

\begin{figure*}
	\centering
	\includegraphics[width=\textwidth]{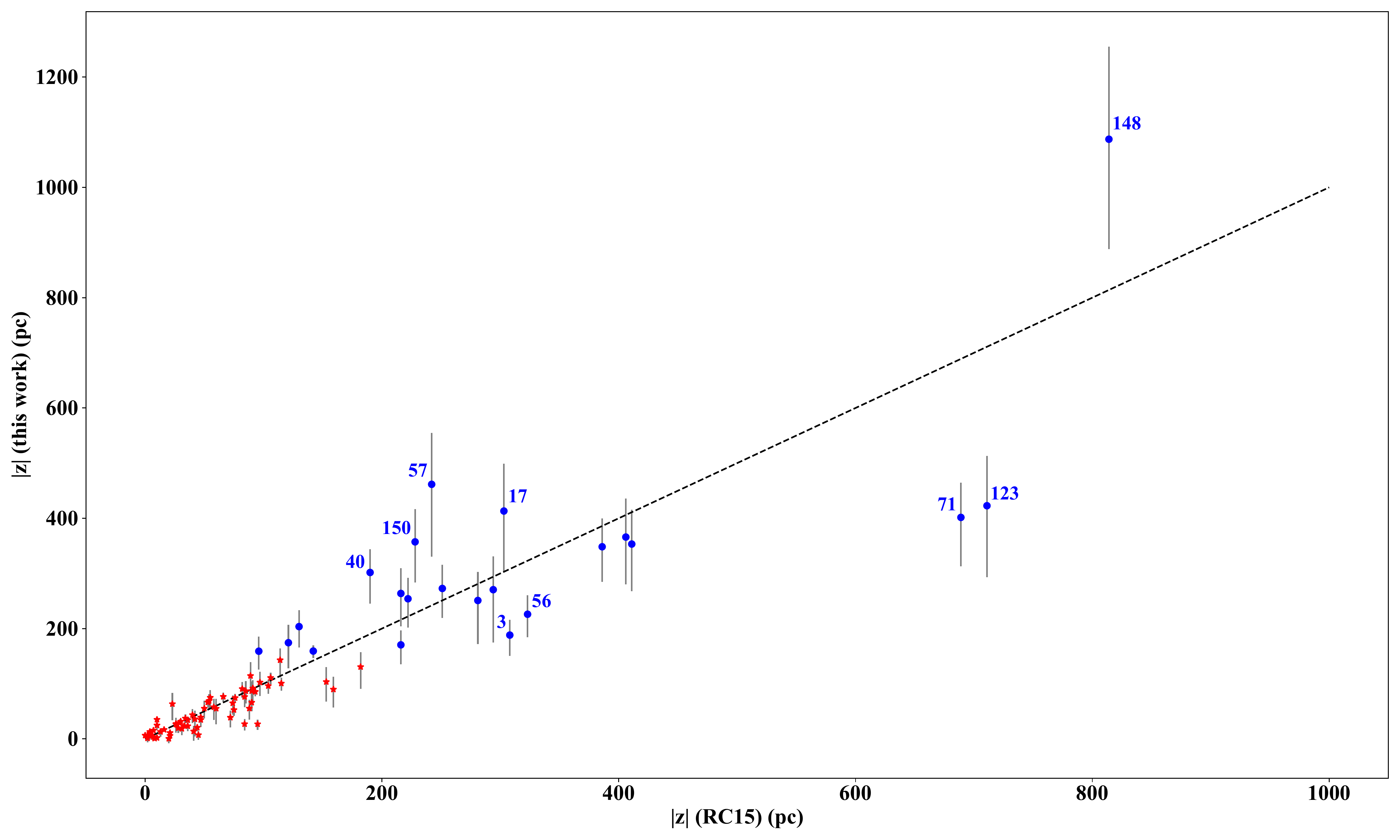}
	\caption{A comparison between the WR distances from the midplane from \citet{2015MNRAS.447.2322R} and this work. Blue circles are the points from this work with distances greater than 3$\sigma$, where $\sigma$ is the \hii region scale height. The dotted line indicates parity between the two measures. Stars with significant disagrement are labelled with their WR numbers.} 
	\label{fig:hi_ori}
\end{figure*}

\subsection{Comparison with previous WR distances} \label{ssec:compold}

\citet{2015MNRAS.447.2322R} provide distance estimates for 228 Galactic WR stars based on previous absolute magnitude calibrations. Of those, 87 have reliable distances from this work.
Fig.~\ref{fig:cf_ori}(a) compares distances to Galactic WR stars in common with \citet{2015MNRAS.447.2322R}. Agreement is reasonable up to $\sim$2 kpc.  This is the subset of \textit{Gaia} sources with the lowest uncertainties and extinction, enabling accurate applications of our prior and absolute magnitude calibrations. Beyond 2 kpc, there is significant scatter, with many stars closer than previously thought. These are principally more highly reddened WR stars that have been discovered recently. Conversely many stars that were thought to be nearby based on calibrations, have significantly larger distances (e.g. WR57 is revised from 2.98$\pm$0.52 kpc to 5.50$\substack{+1.49 \\ -1.06}$ kpc).

All of our 187 stars with reliable absolute magnitudes have distance estimates from \citet{2018AJ....156...58B}. Comparisons are presented in Figure~\ref{fig:cf_ori}(b). Again, good agreement is obtained up to $\sim$2 kpc, beyond which the \citet{2018AJ....156...58B} distances are generally larger than our results. The average $\omega/\sigma_{\omega}$ for stars at distances beyond 2.5 kpc is $-$0.71. The error is therefore a substantial proportion of the total parallax, which suggests disparities stem primarily from limitations in the \textit{Gaia} data and the differences between the two priors. At large distances and so proportionally large parallax errors, the prior dominates the data and the peak of the posterior shifts closer to the peak of the prior. 

For this work, the peak of the prior probability defaults to <3 kpc, depending on longitude. If the peak in the Bailer-Jones prior is substantially closer or further, this results in a large divergence between the two measures. Our prior differs significantly from \citet{2018AJ....156...58B} as it more directly accounts for extinction and the specific distribution of massive stars. The red stars/black crosses in Figure~\ref{fig:cf_ori}(b) show the contrast between results calculated with/without the dust extinction model. In most instances, the stars had results more in line with \citet{2018AJ....156...58B} when dust was excluded. Therefore, in the vast majority of cases, dust extinction in the prior is the primary factor leading to different results.

Since distances from  \citet{2018AJ....156...58B} formed the basis of the recent spectroscopic studies of Galactic WR stars by \citet{2019A&A...621A..92S} and \citet{2019A&A...625A..57H}, use of distances from this study with no warning flags would lead to generally modest 0.05 dex reductions in stellar luminosity. These are included in Table~\ref{table:final}, with higher reductions for relatively distant stars including WR74 (WN7o, 0.24 dex),  WR91 (WN7b, 0.23 dex) ,WR56 (WC7, 0.20 dex) and WR64 (WC7, 0.20 dex).

We also compare the distances to a Galactic LBV (WR31b = AG Car) and LBV candidate (WR31a = He~3-519) which are in common with \citet{2019MNRAS.488.1760S}. They obtain a distance of 7.12$\substack{+2.53 \\ -1.67}$ kpc to WR31a, versus 7.35$\substack{+1.45 \\ -1.18}$ kpc from this work, and 4.65$\substack{+1.43 \\ -0.92}$ kpc to WR31b, versus 4.85$\substack{+0.93 \\ -0.70}$ kpc from this work. These are well within the uncertainties of both stars, particularly given WR31a has a high error to parallax ratio of 0.72 (as measured directly from the catalogue values). \citet{2019MNRAS.488.1760S} adopt a different zero point to our study, namely $-$0.05\,mas as an initial value and model some uncertainty in this as part of their calculation. This decision is based on the variety of different zero points found in the literature (e.g \citealt{2018ApJ...861..126R}, \citealt{2019ApJ...878..136Z}, \citealt{2018ApJ...862...61S} and \citealt{2019ApJ...872...85G}).

Therefore, these distances are systematically closer than those from \citet{2018AJ....156...58B}. This result agrees both with our findings and \citet{2019MNRAS.487.3568S}, who also find that \citet{2018AJ....156...58B}  appear to systematically overestimate distances. As \citet{2019MNRAS.488.1760S} adopts a similar prior to that of \citet{2018AJ....156...58B}, the overlapping results therefore indicate that the larger zero point is performing much the same function as our dust model, acting to moderate the distances of \citet{2018AJ....156...58B}.



\begin{figure}
	\includegraphics[width=\linewidth]{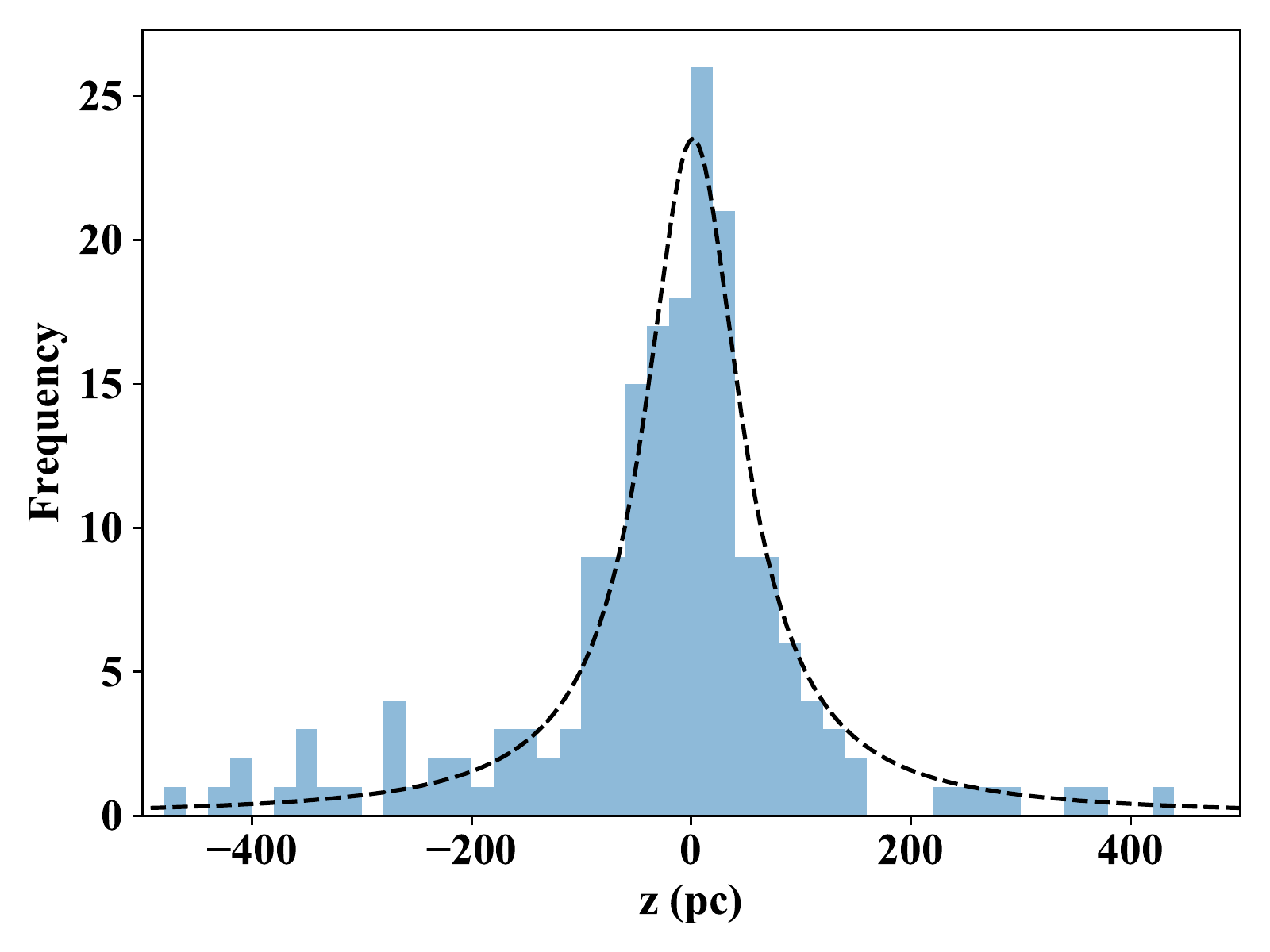}
  \caption{A histogram distribution of WR distances from the Galactic disk. The dotted line shows the Cauchy fit from Equation ~\ref{heightsfit}.} 
	\label{fig:szplt}
\end{figure}

\section{Distances from the Galactic disk} \label{sec:hab}

To identify potential runaway stars, we calculated distances from the Galactic plane using the most likely distance from the Sun and the Galactic latitude of the star, with the addition of the 20.8 pc \citep{2019MNRAS.482.1417B} for the Sun's distance above the midplane. The 68\% distance uncertainty intervals were scaled to give height uncertainties.

The new midplane distances in Table ~\ref{table:final} are compared with results from \citet{2015MNRAS.447.2322R} in Figure ~\ref{fig:hi_ori}. In general, the deviation from previous results increases with height, reflecting the uncertainty of distances to very remote WR stars. The scale heights, $\sigma$, of \hii regions loosely trace massive star formation sites and can therefore highlight potential runaways. Based on the median north scale height between 3.9 kpc and 5.6 kpc in \citet{2004MNRAS.347..237P}, $\sigma$ is 52 pc. The south scale heights contained too few points to be reliable. 

We additionally calculated the scale height of the WR population. The histogram of WR distances from the midplane is presented in Figure ~\ref{fig:szplt} and can be fit with a Cauchy distribution 
\begin{equation} \label{heightsfit}
  g = \frac{A}{\pi\gamma}\frac{\gamma^2}{(z-c)^2+\gamma^2}
\end{equation} 
where A is the scale constant, c is the distribution centre and $\gamma$ is the scale parameter, specifying the half width half maximum (HWHM). Fitting these parameters gives a centre of 1.5 pc and a HWHM of 53.4 pc. The central value of our distribution is similar to \citet{2015MNRAS.447.2322R} (1.9 pc), though their HWHM is somewhat smaller, at 39.2 pc. The central value would suggest many WR stars are slightly above the plane, but this may be due to planar dust extinction rendering WR stars which sit below the disk being inaccessible to \textit{Gaia}.

Our results are similar to \citet{1990AJ....100..431C}, who find a WR scale height of 45$\pm$5 pc using an isothermal disk model and \citet{2016AstL...42....1B}, who obtained a height 51.3$\pm$3.7 pc using the same method. However, this latter value relies on a sample at <4 kpc (excluding distant stars to avoid the effects of Galactic disk warp) and thus only covers about half the WR stars in our sample. 

To identify only the most extreme runaways and ensure they did not form in situ, we apply a 3$\sigma$ cut-off using the \hii region scale height. Since a velocity of 1 km\,s$^{-1}$ equates to 1 pc\,Myr$^{-1}$, runaways ($\geq$30 km\,s$^{-1}$) will travel in excess of 150 pc over a typical WR lifetime of 5 Myr. 91\% of 383 WR stars in \textit{Gaia} reside within three scale heights from the Galactic plane, so 9 \% of WR stars are located far from the Galactic plane. Table ~\ref{table:zheights} presents the |z| distances for each of these stars.

However, the resulting runaway list does not acount for the known warp in the Galactic disk. \citet{2019A&A...627A.150R} estimate the warp begins at a radius of 12-13 kpc from the Galactic centre for their sample of young, bright stars (which they refer to as the OB sample). All but two of our WR stars are within 12 kpc of the Galactic centre and by this measure, would be unaffected. However, their results show some complex structures that in fact suggest some of our sample may be affected by the warp. An alternative measure from \citet{2019ApJ...871..208L}, estimates that the Galactic disk instead begins to warp at a radius of 9.2 kpc. 20 stars are further from the centre than this distance, and so their heights would need to account for the warp.

To obtain a robust candidate list of runaways with $\geq$30 km\,s$^{-1}$, we used the Galactic warp model and onset from \citet{2019ApJ...871..208L} to calculate the height of the Galactic plane at the position of each of the 383 WR stars with distances. We subtracted off the height of this Galactic warp, which produced a distance from the midplane for each star, which accounted for the warp. These distances were then used to exclude any stars which were not 3$\sigma$ from the plane, once the warp was accounted for. Using this method, we excluded WR8 and WR12 from our runaway list in Table \ref{table:zheights}. Therefore, 31 stars (8\% of WR stars in \textit{Gaia}) are robust runaway candidates.

We do not apply the warp to our full list of distances from the plane in Table \ref{table:final}, as the warp onset and model are still uncertain.

The runaways identified in \citet{2015MNRAS.447.2322R} generally remain far from the plane. However, many of the more extreme distances from the plane are now moderated, due to reduced distances from the Sun. This suggests that extreme runaways are less common than previously thought. WR93a and WR64 are not included, as they were identified as having abnormal $\mathrm{v^{WR}}$ band extinction (Section ~\ref{sec:absmag}), which meant it was not possible to calculate their absolute magnitudes, so their distances could not be validated.

\renewcommand{\arraystretch}{1.5}
\begin{table}
  \caption{Distance of WR stars from the midplane |z|, for which excesses exceed 3$\sigma$, where $\sigma$=52 pc, the \hii region scale height of 52 pc. Previously identified runaways with |z| $\geq$300 pc according to \citet{2015MNRAS.447.2322R} are also indicated}.
  \begin{tabular}{|p{.1\linewidth}|p{.19\linewidth}|p{.12\linewidth}|p{.1\linewidth}|p{.08\linewidth}|p{.15\linewidth}}
  \hline
  WR Number & Spectral type & Dist (kpc) & |z| (pc) & \hii $\sigma$ & Known runaway \\
  \hline
  WR148 & WN8h+ & 9.47$\substack{+1.77 \\ -1.49}$ & 1087$\substack{+199 \\ -168}$ & 20.9$\substack{+3.8 \\ -3.2}$ & Yes \\
  WR57 & WC8 & 5.50$\substack{+1.49 \\ -1.06}$ & 462$\substack{+131 \\ -93}$ & 8.9$\substack{+2.5 \\ -1.8}$ & No \\
  WR123 & WN8o & 5.35$\substack{+1.56 \\ -1.09}$ & 423$\substack{+129 \\ -91}$ & 8.1$\substack{+2.5 \\ -1.7}$ & Yes \\
  WR73 & WC9d & 6.81$\substack{+1.85 \\ -1.47}$ & 423$\substack{+109 \\ -87}$ & 8.1$\substack{+2.1 \\ -1.7}$ & No \\
  WR17 & WC5 & 6.75$\substack{+1.74 \\ -1.33}$ & 413$\substack{+112 \\ -86}$ & 7.9$\substack{+2.1 \\ -1.6}$ & Yes \\
  WR71 & WN6o & 3.19$\substack{+0.67 \\ -0.48}$ & 402$\substack{+89 \\ -63}$ & 7.7$\substack{+1.7 \\ -1.2}$ & Yes \\
  WR6 & WN4b & 2.27$\substack{+0.42 \\ -0.31}$ & 376$\substack{+73 \\ -54}$ & 7.2$\substack{+1.4 \\ -1.0}$ & No \\
  WR75c & WC9 & 7.15$\substack{+1.78 \\ -1.45}$ & 366$\substack{+86 \\ -70}$ & 7.0$\substack{+1.7 \\ -1.3}$ & Yes \\
  WR124 & WN8h & 5.87$\substack{+1.48 \\ -1.09}$ & 360$\substack{+85 \\ -63}$ & 6.9$\substack{+1.6 \\ -1.2}$ & Yes \\
  WR150 & WC5 & 8.73$\substack{+1.70 \\ -1.38}$ & 357$\substack{+73 \\ -60}$ & 6.9$\substack{+1.4 \\ -1.1}$ & No \\
  WR61 & WN5o & 5.49$\substack{+1.25 \\ -0.91}$ & 353$\substack{+85 \\ -62}$ & 6.8$\substack{+1.6 \\ -1.2}$ & Yes \\
  WR49 & WN5(h) & 8.35$\substack{+1.44 \\ -1.17}$ & 348$\substack{+64 \\ -52}$ & 6.7$\substack{+1.2 \\ -1.0}$ & Yes \\
  WR58 & WN4b/CE & 5.88$\substack{+1.42 \\ -1.04}$ & 337$\substack{+86 \\ -63}$ & 6.5$\substack{+1.7 \\ -1.2}$ & No \\
  WR40 & WN8h & 3.83$\substack{+0.67 \\ -0.50}$ & 302$\substack{+56 \\ -42}$ & 5.8$\substack{+1.1 \\ -0.8}$ & No \\
  WR126 & WC5/WN & 7.57$\substack{+1.49 \\ -1.19}$ & 300$\substack{+55 \\ -44}$ & 5.8$\substack{+1.1 \\ -0.8}$ & No \\
  WR103 & WC9d+? & 3.46$\substack{+1.28 \\ -0.77}$ & 275$\substack{+109 \\ -65}$ & 5.3$\substack{+2.1 \\ -1.3}$ & No \\
  WR33 & WC5; WC6 & 7.59$\substack{+1.62 \\ -1.30}$ & 273$\substack{+54 \\ -43}$ & 5.2$\substack{+1.0 \\ -0.8}$ & No \\
  WR69 & WC9d+OB & 3.48$\substack{+0.64 \\ -0.47}$ & 272$\substack{+54 \\ -40}$ & 5.2$\substack{+1.0 \\ -0.8}$ & No \\
  WR92 & WC9 & 3.78$\substack{+1.25 \\ -0.79}$ & 271$\substack{+96 \\ -61}$ & 5.2$\substack{+1.8 \\ -1.2}$ & No \\
  WR54 & WN5o & 6.52$\substack{+1.37 \\ -1.05}$ & 264$\substack{+60 \\ -46}$ & 5.1$\substack{+1.1 \\ -0.9}$ & Yes \\
  WR129 & WN4o & 5.47$\substack{+1.22 \\ -0.90}$ & 254$\substack{+52 \\ -38}$ & 4.9$\substack{+1.0 \\ -0.7}$ & No \\
  WR83 & WN5o & 3.80$\substack{+1.10 \\ -0.72}$ & 251$\substack{+79 \\ -52}$ & 4.8$\substack{+1.5 \\ -1.0}$ & No \\
  WR131 & WN7h+abs & 6.92$\substack{+1.40 \\ -1.09}$ & 227$\substack{+42 \\ -32}$ & 4.4$\substack{+0.8 \\ -0.6}$ & No \\
  WR56 & WC7 & 8.67$\substack{+1.46 \\ -1.20}$ & 226$\substack{+41 \\ -34}$ & 4.3$\substack{+0.8 \\ -0.7}$ & Yes \\
  WR30 & WC6+O6-8 & 5.09$\substack{+0.99 \\ -0.74}$ & 211$\substack{+45 \\ -33}$ & 4.1$\substack{+0.9 \\ -0.6}$ & No \\
  WR20 & WN5o & 6.98$\substack{+1.18 \\ -0.93}$ & 204$\substack{+38 \\ -30}$ & 3.9$\substack{+0.7 \\ -0.6}$ & No \\
  WR3 & WN3ha & 2.90$\substack{+0.52 \\ -0.39}$ & 188$\substack{+38 \\ -28}$ & 3.6$\substack{+0.7 \\ -0.5}$ & Yes \\
  WR4 & WC5+? & 3.75$\substack{+0.89 \\ -0.62}$ & 174$\substack{+47 \\ -32}$ & 3.4$\substack{+0.9 \\ -0.6}$ & No \\
  WR128 & WN4(h) & 2.90$\substack{+0.54 \\ -0.39}$ & 170$\substack{+35 \\ -26}$ & 3.3$\substack{+0.7 \\ -0.5}$ & No \\
  WR52 & WC4 & 1.75$\substack{+0.16 \\ -0.13}$ & 159$\substack{+13 \\ -11}$ & 3.1$\substack{+0.2 \\ -0.2}$ & No \\
  WR34 & WN5o & 7.41$\substack{+1.37 \\ -1.09}$ & 159$\substack{+33 \\ -26}$ & 3.1$\substack{+0.6 \\ -0.5}$ & No \\
  \hline
  \end{tabular}
  \label{table:zheights}
\end{table}

Two main evolutionary paths may have created these runaways. The first is the disruption of a binary system when the primary star explodes as a supernova and ejects the remaining companion \citep{1961BAN....15..265B}. The second scenario is dynamical ejection from a dense cluster, which can eject both binary and single stars \citep{1967BOTT....4...86P}. The majority of outliers with >3$\sigma$ distances are apparently single stars, as only WR30 and WR69 have confirmed OB companions.

As both single stars and binaries can be ejected from clusters, it is not possible for us to definitively state which mechanism is dominant. We defer a discussion of the origin of runaways to Paper II which considers the association of WR stars with star clusters or OB associations. However, we note that recent simulations suggest fast runaways from either mechanism are anticipated to be very rare \citep{2019A&A...624A..66R, 2016A&A...590A.107O}, in stark contrast with the high fraction of WR stars at extreme distances from the Galactic plane. 


Two stars merit individual consideration. The high velocity runaway WR124 is now located at |z|=360 pc, compared to previous estimates of 217 pc \citep{2015MNRAS.447.2322R}, 193 pc \citep{2010ApJ...724L..90M} and 250 pc \citep{1982A&A...114..135M}. This confirms its runaway status, although our work places it significantly further from the Sun (5.9 kpc instead of 3.3 kpc from \citealt{2010ApJ...724L..90M}). 

WR148 is located furthest from the Galactic plane. \citet{1986ApJ...304..188D} suggested it as a possible WR+compact object binary disrupted by a SN, however, \citet{2017MNRAS.467.3105M} claim it is instead a WN+O binary. If the latter is true, our data suggests that WR148 is a binary system that has been ejected from a cluster, concurring with \citet{2017MNRAS.467.3105M}. Assuming a lifetime of 5 Myr and a straight vertical trajectory from the Galactic disk, the minimum possible velocity for WR148 is 212 \kms, making it a very rapid cluster ejection.

\citet{1989ApJ...347..373M} suggested WN8-9 were over represented amongst runaways, a finding which was corroborated by \citet{2015MNRAS.447.2322R}. However amongst our sample, only 4 out of 31 stars are of the WN8-9 subtype. The previous over representation disappears with the drop in extreme runaways. If our sample is representative of the wider WR star population, this suggests that the observed distribution was due to overestimated distance measurements, which would have made the stars appear further from the plane than they truly are. 


\section{Conclusions} \label{sec:con}

We have calculated distances and absolute magnitudes of the Galactic WR population using data from \textit{Gaia} DR2:

\begin{itemize}[leftmargin=*]
  
  \item 383 WR stars (58\% of the known Galactic population) have full five parameter astrometric solutions (proper motions and parallaxes) in the \textit{Gaia} catalogue. WR stars with large J$-$K>3 colours, indicating high dust extinctions, were generally not detected by \textit{Gaia}.

  \item We used the \textit{Gaia} parallaxes to calculate distances to the 383 WR stars detected by \textit{Gaia}. We use Bayesian methods to properly transform the parallax uncertainties to distance uncertainties and to obtain distances from negative parallaxes. Our Bayesian prior accounts for extinction using a Galactic dust model and the specific distribution of massive stars using \hii regions. Potential underestimates of parallax uncertainties and the zero point error are accounted for in our calculation. 

  \item The resulting distances agree well with both the previous calibration \citep{2015MNRAS.447.2322R} and DR2 distances from \citet{2018AJ....156...58B} up to 2 kpc. Deviations above 2 kpc are due primarily to the large uncertainties of the \textit{Gaia} parallaxes.  Distances from  \citet{2018AJ....156...58B} formed the basis of recent spectroscopic studies of Galactic WR stars by \citet{2019A&A...621A..92S} and \citet{2019A&A...625A..57H}. Use of distances from this study would generally lead to modest 0.05 dex reductions in stellar luminosities, albeit with reductions of up to 0.2 dex for relatively distant stars.
  
  \item 25 WR stars are found within 2 kpc, compared to 30 WR stars from \citet{1983ApJ...274..302C}. Based on the population in GOSC v4.1 \citep{2013msao.confE.198M}, the WR/O star ratio in this region is 0.09.

  \item We calculate absolute magnitudes for WR stars, in both the $\mathrm{v^{WR}}$ and $\mathrm{K_s}$ bands. Of these, 187 stars have an absolute magnitude in either band and were used to generate subtype averages for calibrations. Both WN and WC stars are found to be more diverse in their absolute magnitude ranges than anticipated and therefore we recommend avoiding use of calibrations without accounting for this large intrinsic spread. 

  \item We have applied our new distances to identify 31 potential runaways from the Galactic disk, accounting for the Galactic warp. \hii region scale heights define the cut-offs for runaway status. 20 of these WR stars with |z|>156 pc are new detections. The vast majority of the runaway stars are single. However, as both companion supernovae and dynamical ejection from clusters can produce single star runaways, it was not possible for us to determine the dominant runaway production mechanism, which is deferred to Paper II.
  
\end{itemize}

The current limitations of our prior are mainly the simplified dust extinction map. With an increased number of observations, the quality of future \textit{Gaia} release data should improve. Therefore, the number of WR stars with negative parallaxes should fall and we thus expect a corresponding decrease in the number of flagged results. Better parallax to error ratios in the early DR3 release (estimated to improve by a factor 1.2, \citealt{gaia_pres2}), will also reduce uncertainties and the effect of our prior when used with small parallaxes. Further improvements to the astrometric modelling and fitting algorithms should also reduce the number of questionable results via a reduction in astrometric excess noise. Finally, there is a possibility that the number of WR stars with distances will increase. 32 objects only had two parameter solutions (fitting positions) from \textit{Gaia} DR2. Future \textit{Gaia} data releases may find satisfactory full five parameter solutions, which would also include parallaxes.


\section*{Acknowledgements}

GR wishes to thank the Science and Technology Facilities Council (STFC), for their financial support through the Doctoral Training Partnership.

We wish to thank the referee Dr Anthony Brown for his helpful comments and suggestions on the submitted manuscript. We also thank Dr Josep Manel Carrasco and Dr Carme Jordi for providing the synthetic photometry in V broadband, \textit{Gaia} $G_{BP}-G_{RP}$ and G filters at different extinctions and for different WR star subtypes, used in Section ~\ref{ssec:extinctions}. 

This work has made use of data from the European Space Agency (ESA) mission {\it \textit{Gaia}} (\url{https://www.cosmos.esa.int/gaia}), processed by the {\it \textit{Gaia}} Data Processing and Analysis Consortium (DPAC, \url{https://www.cosmos.esa.int/web/gaia/dpac/consortium}). Funding for the DPAC has been provided by national institutions, in particular the institutions participating in the {\it \textit{Gaia}} Multilateral Agreement.

This publication also makes use of data products from the Two Micron All Sky Survey, which is a joint project of the University of Massachusetts and the Infrared Processing and Analysis Center/California Institute of Technology, funded by the National Aeronautics and Space Administration and the National Science Foundation.

The work in Section ~\ref{ssec:gcat} is based on data products from observations made with ESO Telescopes at the La Silla Paranal Observatory under programme ID 177.D-3023, as part of the VST Photometric H$\mathrm{\alpha}$ Survey of the Southern Galactic Plane and Bulge (VPHAS+, www.vphas.eu). Additionally, this paper makes use of data obtained as part of the INT Photometric H$\mathrm{\alpha}$ Survey of the Northern Galactic Plane (IPHAS, www.iphas.org) carried out at the Isaac Newton Telescope (INT). The INT is operated on the island of La Palma by the Isaac Newton Group in the Spanish Observatorio del Roque de los Muchachos of the Instituto de Astrofisica de Canarias. All IPHAS data are processed by the Cambridge Astronomical Survey Unit, at the Institute of Astronomy in Cambridge. The bandmerged DR2 catalogue was assembled at the Centre for Astrophysics Research, University of Hertfordshire, supported by STFC grant ST/J001333/1.

In addition to Astropy \citep{2013A&A...558A..33A}, this work would not be possible without the python packages Numpy (\citealt{numpy_book}, \citealt{doi:10.1109/MCSE.2011.37}), Pandas \citep{mckinney-proc-scipy-2010} and Matplotlib \citep{2007CSE.....9...90H}.


\bibliographystyle{mnras}


\bibliography{gaia_dr2_pc} 
\makeatletter
\relax
\def\mn@urlcharsother{\let\do\@makeother \do\$\do\&\do\#\do\^\do\_\do\%\do\~}
\def\mn@doi{\begingroup\mn@urlcharsother \@ifnextchar [ {\mn@doi@}
  {\mn@doi@[]}}
\def\mn@doi@[#1]#2{\def\@tempa{#1}\ifx\@tempa\@empty \href
  {http://dx.doi.org/#2} {doi:#2}\else \href {http://dx.doi.org/#2} {#1}\fi
  \endgroup}
\def\mn@eprint#1#2{\mn@eprint@#1:#2::\@nil}
\def\mn@eprint@arXiv#1{\href {http://arxiv.org/abs/#1} {{\tt arXiv:#1}}}
\def\mn@eprint@dblp#1{\href {http://dblp.uni-trier.de/rec/bibtex/#1.xml}
  {dblp:#1}}
\def\mn@eprint@#1:#2:#3:#4\@nil{\def\@tempa {#1}\def\@tempb {#2}\def\@tempc
  {#3}\ifx \@tempc \@empty \let \@tempc \@tempb \let \@tempb \@tempa \fi \ifx
  \@tempb \@empty \def\@tempb {arXiv}\fi \@ifundefined
  {mn@eprint@\@tempb}{\@tempb:\@tempc}{\expandafter \expandafter \csname
  mn@eprint@\@tempb\endcsname \expandafter{\@tempc}}}


\makeatother



\bsp	
\label{lastpage}
\end{document}


\label{firstpage}
\pagerange{\pageref{firstpage}--\pageref{lastpage}}

\appendix
\section{ADQL query} \label{sec:adquert}

{\fontfamily{qcr}\selectfont{SELECT TOP 10 DISTANCE(POINT('ICRS', ra, dec), POINT('ICRS', WRra, WRdec)) AS dist, * \\
FROM gaiadr2.gaia\_source \\
WHERE CONTAINS(POINT('ICRS', ra, dec), CIRCLE('ICRS', WRra, WRdec, search\_radius))=1 \\
ORDER BY dist ASC \\}}

where WRra and WRdec are the WR RA and DEC search coordinates in decimal format and the search\_radius is one arcsecond. The query selects the top ten closest points (arranged in distance order) that are within a 1'' circle of the WR search coordinates. All \textit{Gaia} catalogue columns are selected for convenience. 

\section{Increased uncertainties} \label{sec:bcert}

Figure 3 in Section 2.2.1 shows the underestimation of DR2 parallax uncertainties, as compared to the uncertainties of external data (from table 1. in \citealt{2018A&A...616A..17A}). The combined Gaussian and straight line fit to the uncertainties is given by:

\begin{equation} \label{eq:uwu}
X = -0.01319 G + 1.376 + \frac{1.1}{\sqrt{2\pi}1.35}\exp\Bigg[-\frac{1}{2(1.35)^2}\bigg(G-14.59\bigg)^2\Bigg]
\end{equation} 

where $G$ is the WR \textit{Gaia} G band magnitude and $X$ is the factor by which the error is estimated to increase. The updated parallax (in mas) $\omega$ and error $\sigma_{\omega}$ (also in mas) parallax inputs to the likelihood are therefore given by

\begin{equation} \label{eq:newpar}
\omega = \Psi+0.029
\end{equation}

\begin{equation} \label{eq:newer}
  \sigma_{\omega} = \sigma_{\Psi}X
\end{equation}  

where $\Psi$ is the original parallax from the \textit{Gaia} catalogue. This leads to a final likelihood of the form

\begin{equation} \label{eq:likelihood}
P(\omega|r,\sigma_{\omega})=\frac{1}{\sqrt{2\pi}\sigma_{\omega}}\exp\Bigg[-\frac{1}{2\sigma_{\omega}^2}\bigg(\omega-\frac{1}{r}\bigg)^2\Bigg]
\end{equation}

\section{Prior details} \label{sec:bprior}

\begin{figure*}
  \centering
  \vspace{-3cm}
  \begin{adjustwidth}{-1.2cm}{1.9cm}
  \setlength{\subfigcapskip}{10pt}
  \subfigure[]{{\includegraphics[scale=0.45]{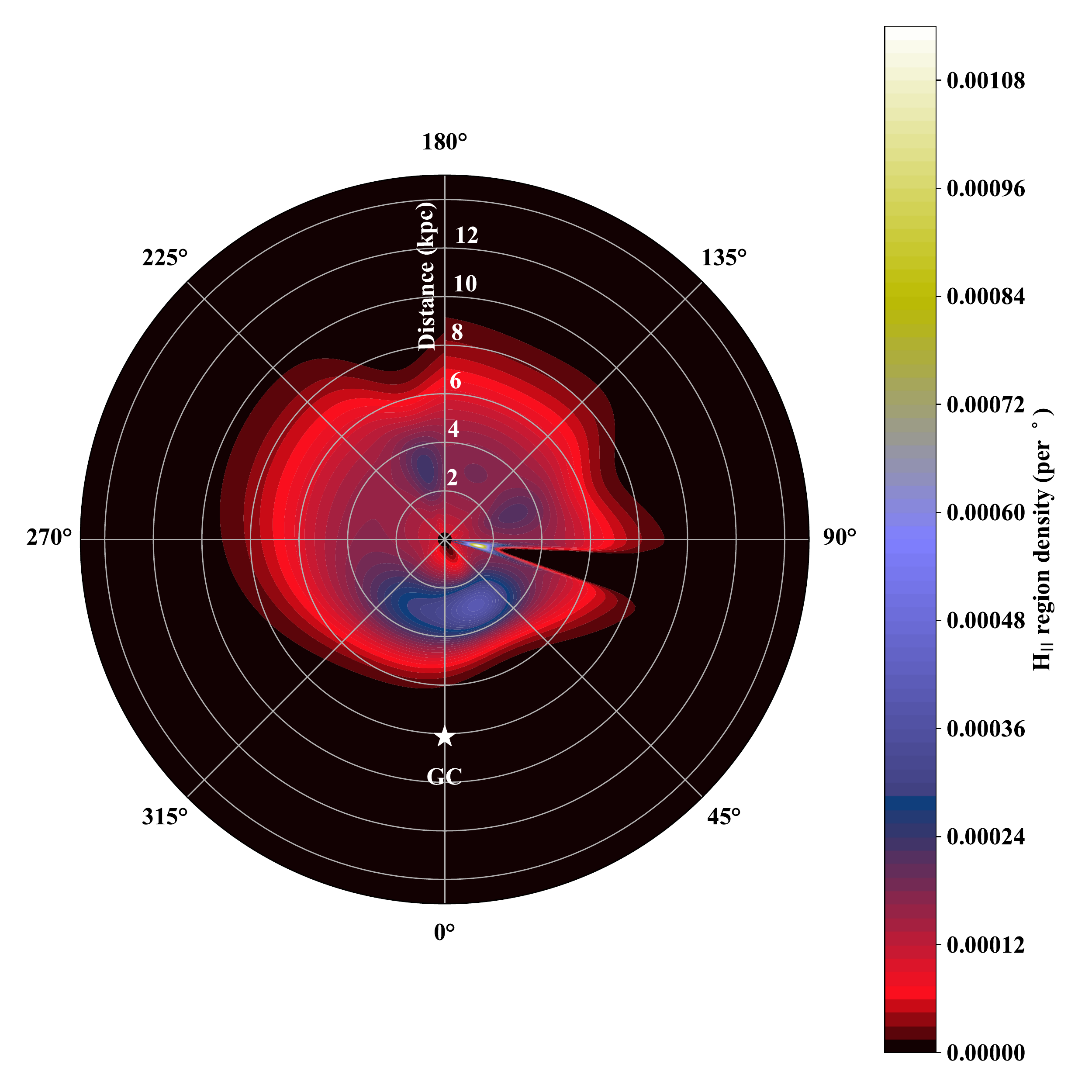}}}
  \hspace{0cm}
  \subfigure[]{{\includegraphics[scale=0.45]{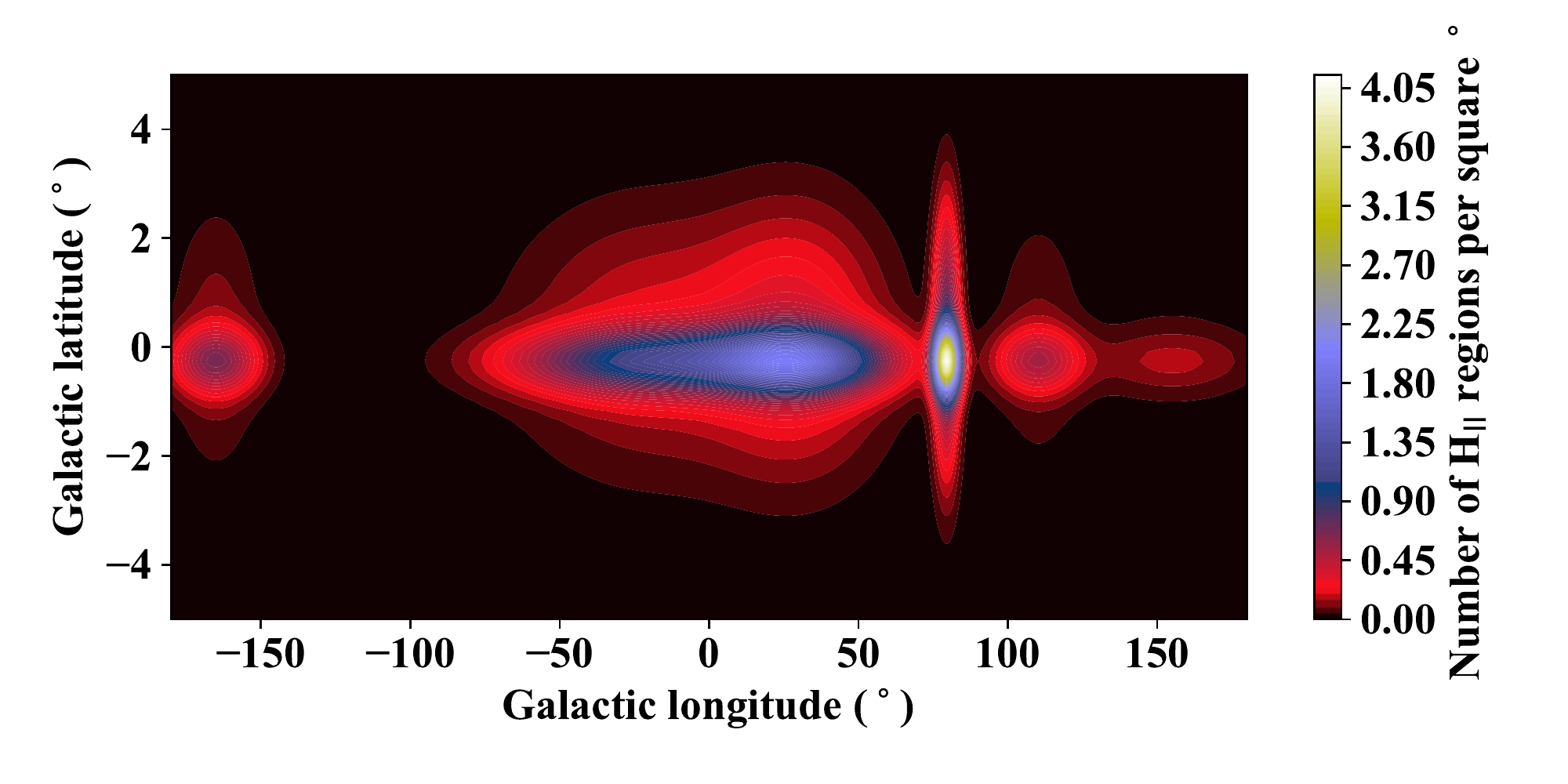}}}
  \end{adjustwidth}  
  \caption{(a) Density of Galactic \hii regions over distance and longitude, at  zero latitude, before extinction is applied (based on \citealt{2004MNRAS.347..237P} and \citealt{2003A&A...397..213P}). The coordinate system is centred on the Sun, with the Galactic Centre at 8.122 kpc. (b) Density of Galactic \hii regions across different latitudes, viewed from the Sun and based on \citet{2003A&A...397..213P}.}
  \label{fig:hii_dist}
\end{figure*}

\begin{figure*}
  \centering
  \begin{adjustwidth}{-1.5cm}{1.7cm}
  \setlength{\subfigcapskip}{10pt}
  \subfigure[]{{\includegraphics[scale=0.45]{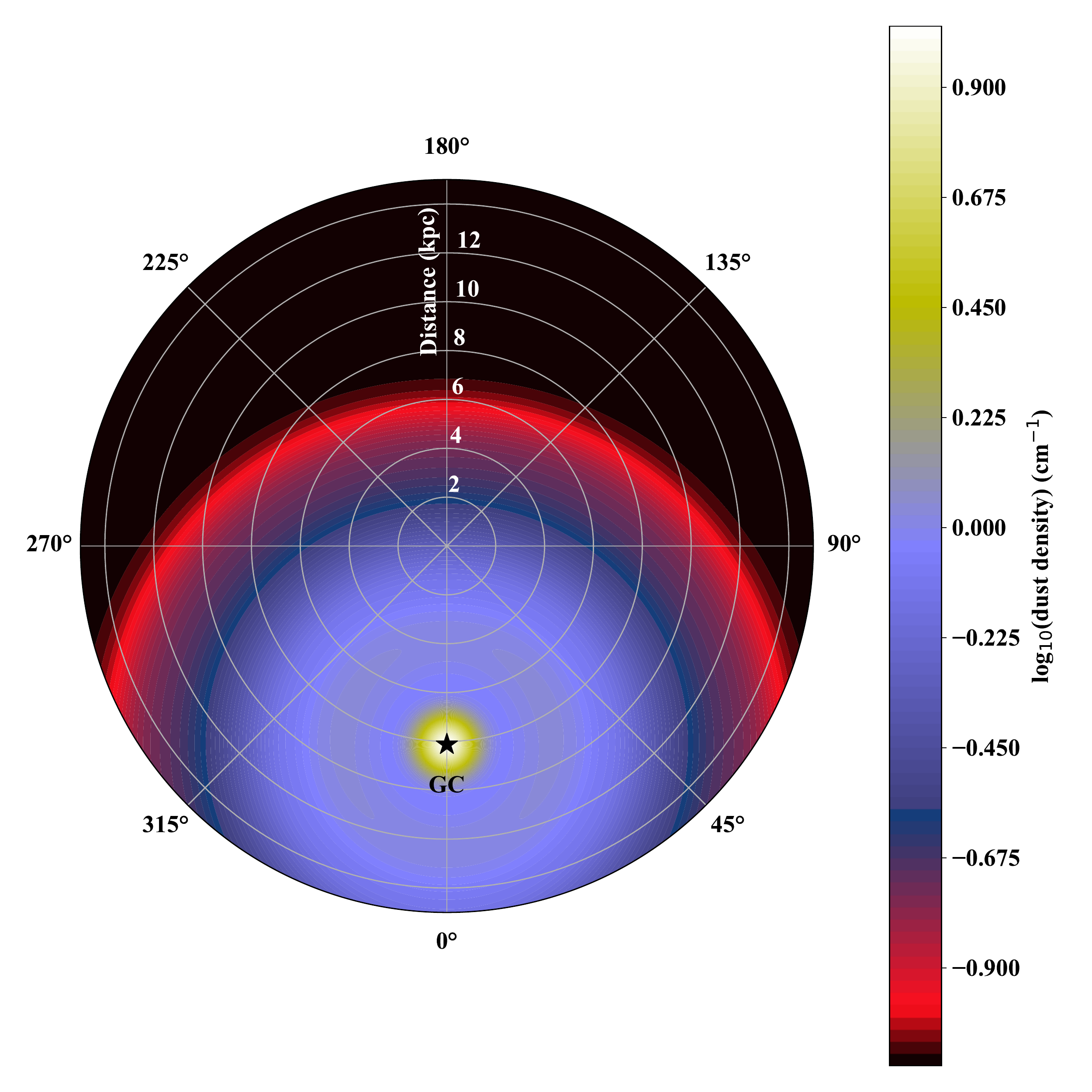}}}
  \hspace{0cm}
  \subfigure[]{{\includegraphics[scale=0.5]{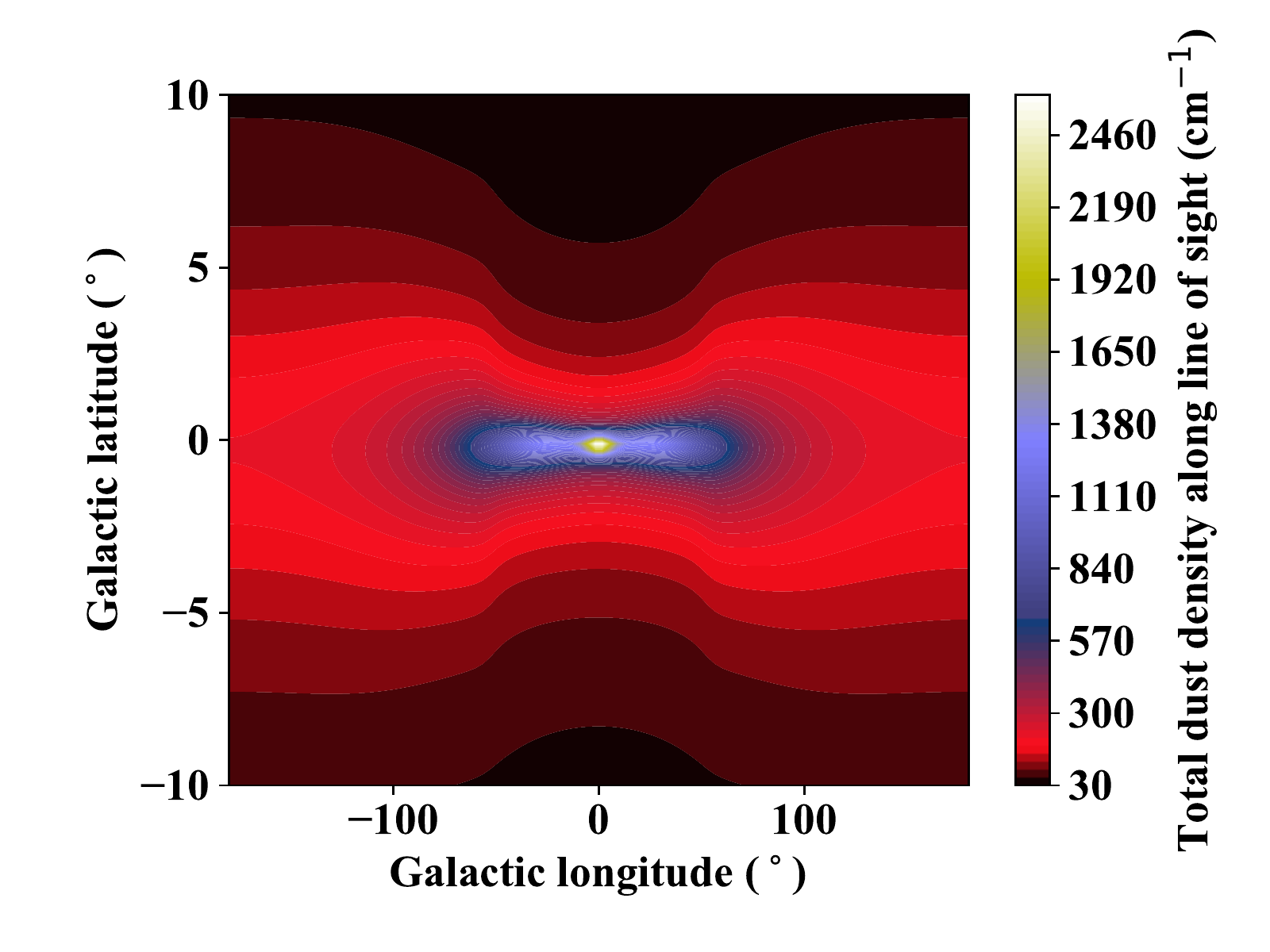}}}
  \end{adjustwidth}  
  \caption{(a) Dust distribution over longitude and distance, at  zero latitude, in the simple disk model and (b) the variation of dust integrated along line of sight with latitude, viewed from the Sun. The coordinate system is centred on the Sun, with the Galactic Centre at 8.122 kpc.}
  \label{fig:dusts}
\end{figure*}

\begin{figure*}
  \centering
  \vspace{-3cm}
	\includegraphics[width=0.7\linewidth]{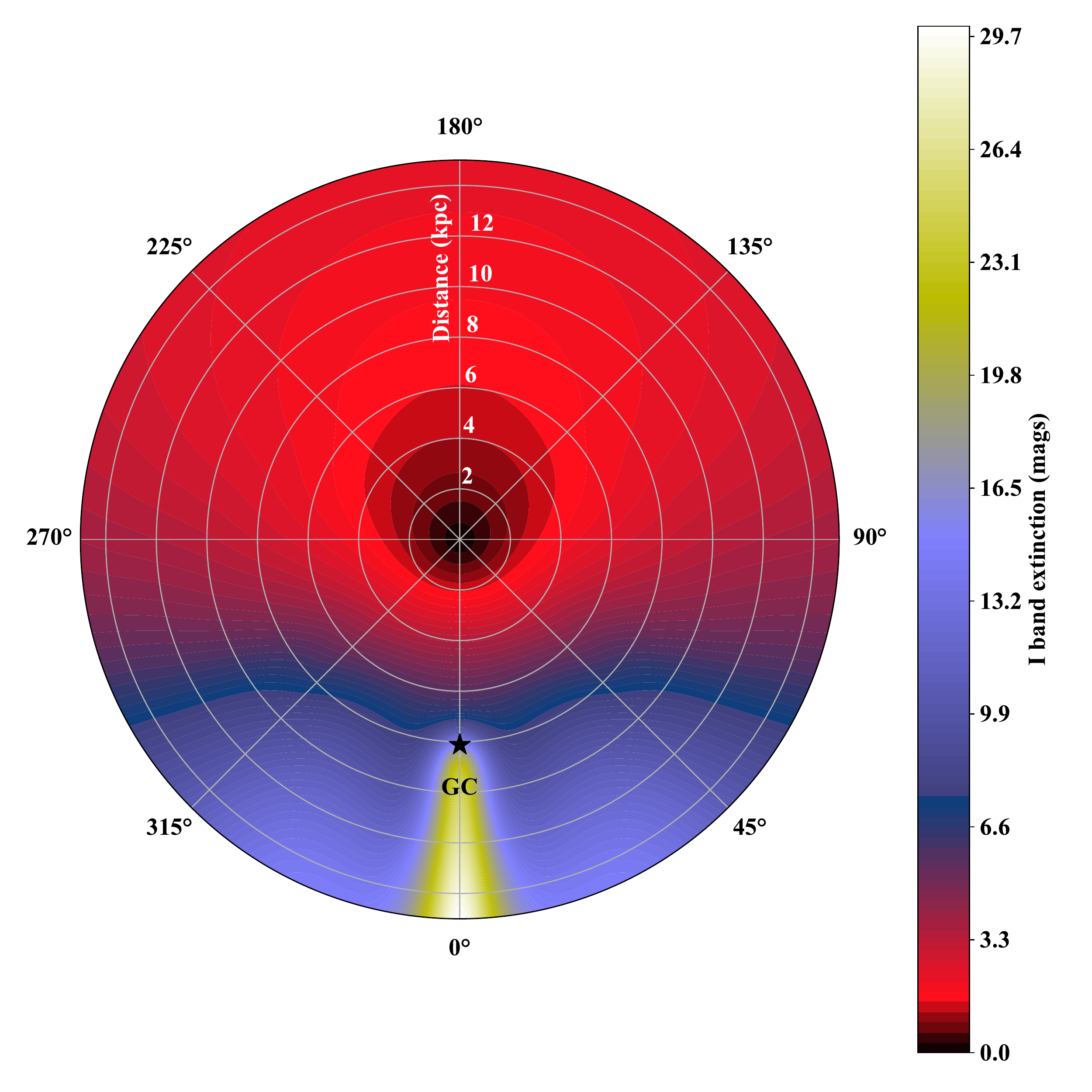} 
	\caption{Extinction variation with distance and Galactic longitude, at  zero latitude, as calculated using the dust model. The plot is
  centred on the Sun, with the Galactic Centre at 8.122 kpc.} 
    \label{fig:ai_long}
\end{figure*}

\begin{figure*}
  \centering
	\includegraphics[width=0.7\linewidth]{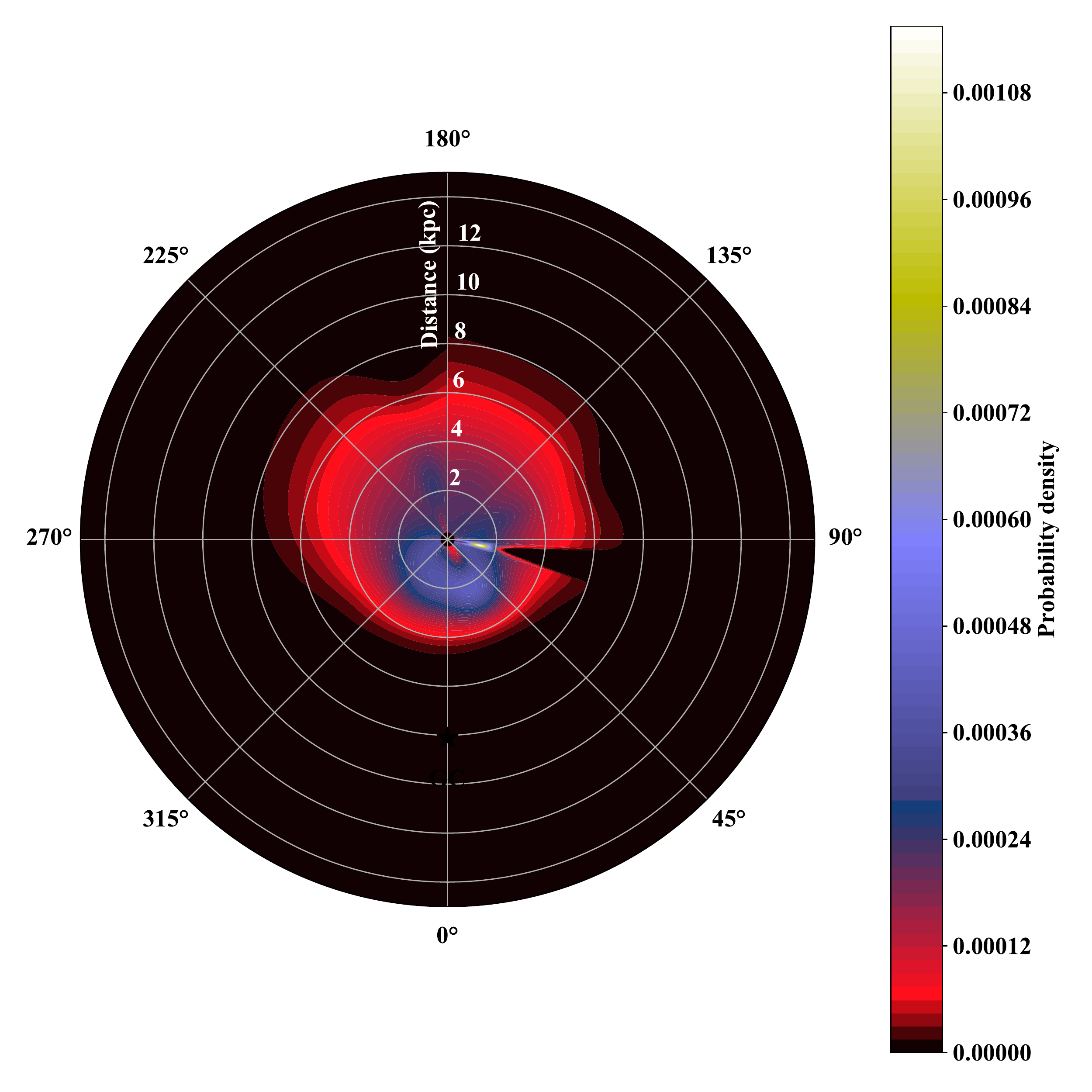} 
  \caption{Combined prior, consisting of \hii region prior and dust extinction.}
    \label{fig:full_prior}
\end{figure*}

The prior is a combination of \hii region distributions at radio wavelengths (to avoid extinction and get a good spatial sample) and an extinction map.

For the \hii region distribution, we chose a Gaussian centred on 3000 pc to best approximate the distribution with distance from the Sun (based on Figure 12 from \citealt{2004MNRAS.347..237P}). Over varying Galactic longitudes and latitudes, the number of \hii regions and their spread over distance changes. To alter the distribution for different lines of sight, the standard deviation was modified based on the \hii region number density at a given latitude and longitude. The standard deviations range from 1-3 kpc, depending on the longitude and latitude of the line of sight. Figure~\ref{fig:hii_dist} (a) shows the resulting distribution over different longitudes at different distances and \ref{fig:hii_dist} (b) shows the distribution over latitudes. 

There is a particularly large excess probability around l=73-86$^{\circ}$ and -3$\leq$b$\leq$-4$^{\circ}$ due to the Cygnus X region (as stated in \citealt{2003A&A...397..213P}). Over these coordinates, the mean of the Gaussian is instead centred on 1400 pc and the standard deviation is correspondingly lower. 

The overhead extinction map in Figure~\ref{fig:ai_long} was generated using a simple dust disk model (see Figure~\ref{fig:dusts} (a) for the variation in longitude and \ref{fig:dusts} (b) for the latitude).

Our primary goal was to determine how extinction affected the observable distances along each line of sight. In regions of high extinction, the peak of the prior would have to be shifted towards the Sun, as the probability of seeing a WR star at a greater distance would decrease. The I band (which peaks at $\sim$8000\ang) is best suited for this, as it operates towards the extreme red end of the \textit{Gaia} G band (at 10500\ang). Any distance that is too faint to observe in this wavelength range would therefore be very faint in G and have a small or nil probability of hosting a WR star that is visible to \textit{Gaia}. At each distance, the dust was integrated along the line of sight and normalised to the extinction at the Galactic centre. This was chosen to be 15.36 magnitudes, in the I band.

Unfortunately, it was not possible to reliably convert $A_I$ to $A_G$, as the conversion relationship given in \citet{2018A&A...616A...4E} does not extend to the large values of $V-I_c$ at the Galactic centre. 

Galactic centre extinction in the I band was calculated by assuming the V band extinction at the same point is 32 mag (based on averaging optical extinction at 0.55$\mu$ from \citealt{2011ApJ...737...73F}) and multiplying by 0.48 \citet{1989ApJ...345..245C} to account for the difference in reddening. Figure ~\ref{fig:ai_long} shows the resulting extinction variation with Galactic longitude.

We then converted the extinction to a factor which could be applied to the probability at each distance, to simulate the reduction of flux from extinction

\begin{equation} \label{eq:dust_ext}
  \delta = 2.512^{(-A_I)}
\end{equation} 

where $A_I$ is the I band extinction at that distance, calculated from $A_I=0.48AV$ (where $AV$ is the V band extinction).

This conversion factor was then multiplied by the \hii region distribution, to give a final distribution. This combines both the radio \hii region observations and dust extinction, and so approximates what might be seen by \textit{Gaia}. This final distribution is shown in Figure \ref{fig:full_prior}. As compared to Figure \ref{fig:hii_dist}, the peak of the prior has moved significantly closer to the Sun (within 1-3 kpc, depending on longitude).

\section{Posterior} \label{sec:bmath}

The Bayesian inferred distribution of distances (the posterior $P(r|\Psi,\sigma_{\Psi})$) is calculated using

\begin{equation} \label{eq:bayes}
  P(r|\Psi,\sigma_{\Psi})=\frac{1}{Z}P(\Psi|r,\sigma_{\Psi})P(r)
  \end{equation} 
  
\citep{2015PASP..127..994B}, where $P(\Psi|r,\sigma_{\Psi})$ is the likelihood (the probability distribution of measured parallaxes, $P(r)$ is the prior (the expected distribution of the distances) and Z is a normalisation constant. 

For our likelihood and prior, the resulting posterior distribution is

\begin{equation} \label{eq:posterior}
  P(\omega|r,\sigma_{\omega})=\frac{1}{\sqrt{2\pi}\sigma_{\omega}\sigma_p}\exp\Bigg[-\frac{1}{2}\Big(\frac{\bigg(\omega-\frac{1}{r}\bigg)^2}{\sigma_{\omega}^2}+\frac{(r-\mu_p)^2}{\sigma_p^2}\Big)\Bigg]\delta
\end{equation} 

where $\sigma_p$ is the standard deviation of the Gaussian from the \hii region prior in the direction of the WR and $\mu_p$ is the mean. We do not account for errors in the WR position, as these are insignificant compared to the simplifications in the prior (such as the simplification of the dust distribution).

We calculated the credible intervals (uncertainties), by cycling through each of the calculated probabilities, beginning with the maximum. At each probability, the corresponding distances either side of the distribution peak were selected. The area under the curve for this distance range could then be compared to the target area (e.g 68\% for one sigma uncertainties). The process was repeated until the area integrated reached or exceeded the required credible interval. This method could also be applied to the two sigma uncertainties (95\%)

Due to the finite nature of the grid, slight deviations from the specified 68\% area occurred, the largest of which was for WR11 (which reached 68.5\% of the area). However, these deviations led to typical interval changes of a few pc or less, below the reasonable precision of our distance calculation.  

\section{Impact of uncertainties} \label{sec:undisc}

\begin{figure}
	\centering
	\includegraphics[width=\linewidth]{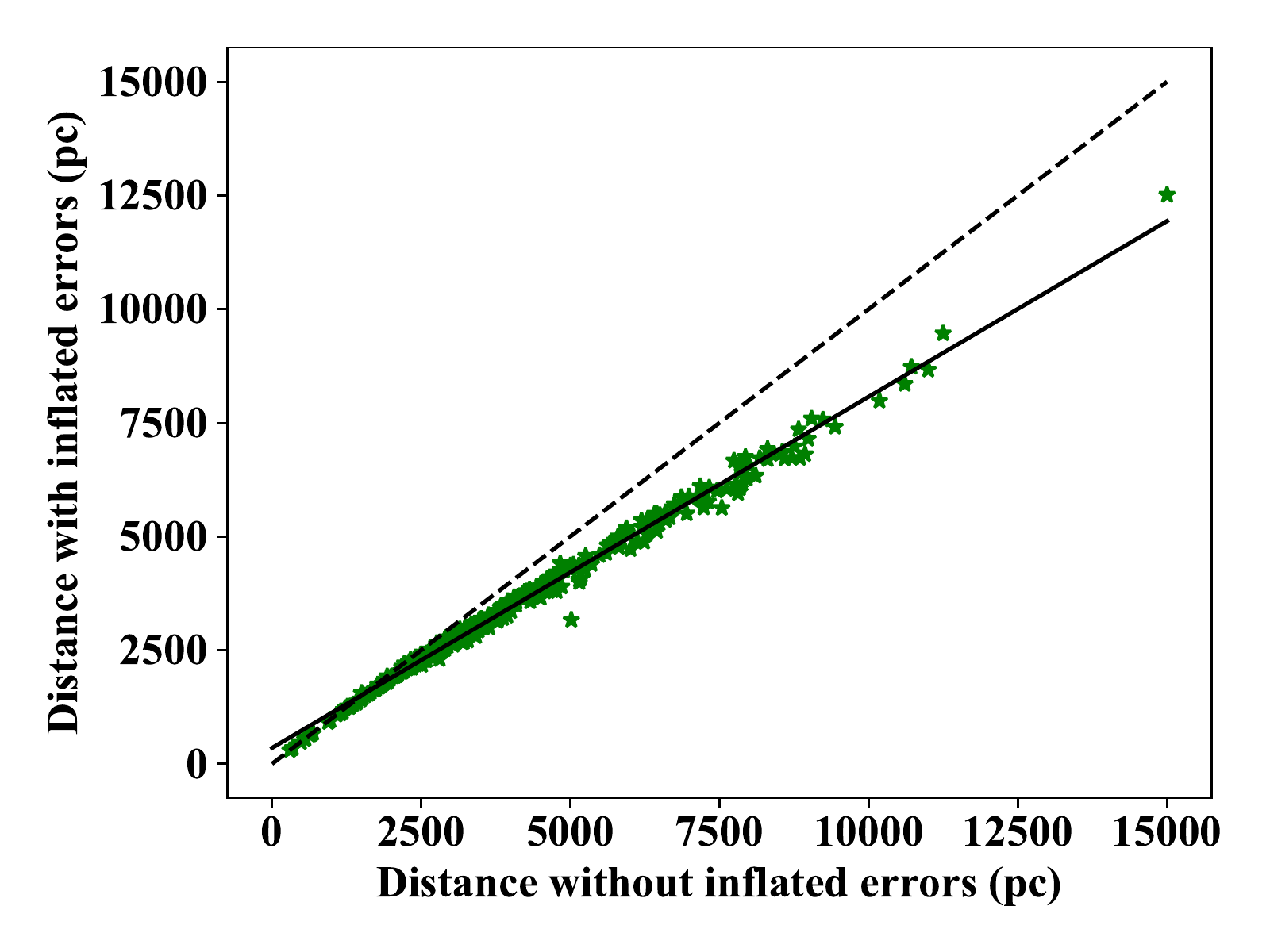} 
	\caption{A comparison between distances with and without the modelled error increase. The dashed line denotes where the two distance calculations are the same and the solid line is the fit from equation \ref{excessfit}} 
    \label{fig:err_dev}
\end{figure}

Figure ~\ref{fig:err_dev} shows that underestimated parallax errors from \textit{Gaia} have a significant effect on the most probable distance. Beyond $\sim$1.4 kpc, the adjusted errors result in systematically closer distances, compared to data with no uncertainty increases. This occurs because the larger parallax to error ratio means the prior has a greater influence on the resulting distance. As the prior accounts for dust extinction, these distances from parallaxes with inflated errors are smaller than those from the original parallaxes.

We can fit a line to determine the typical contribution of modified errors

\begin{equation} \label{excessfit}
  d_e = 0.7724d + 349.25
\end{equation}

where $d_e$ is the distance with extended errors and $d$ contains no error modification. The deviations between this fit and a line x=y indicate a typical contribution of 24\% at 10 kpc, decreasing towards zero at 1.5 kpc. Below this distance, the difference begins to increase again because the increased errors have little effect and the fit is no longer accurate. For isolated cases, the maximum deviation was higher, up to $\sim$50\%.

In most instances, the differences between the distances from the original \textit{Gaia} catalogue parallax error and the distances from the increased parallax error, fall within uncertainties. A major limitation is that the error rescaling used here, may not account for individual errors which are still underestimated. 

Overall, the data show that underestimated parallax errors have a significant effect on many distances and that these underestimates need to be accounted for in distance calculations. 

\section{Bootstrapping and fits to absolute magnitudes} \label{sec:fitdist}

For the bootstrapping procedure, we sample 1000 distributions of 20,000 points each (with replacement) from the true distributions of apparent magnitudes (assumed to be a Gaussian with the peak at the measured value and the standard deviation as the uncertainty), distances and extinction. This generated a distribution of absolute magnitudes which could be fitted with a Gaussian if the $\chi^2$ value was below 0.005 (setting the limit below this value made it difficult to fit stars). Alternatively, if the $\chi^2$ value was above 0.005, a Weibull distribution (non symmetric with left or right skew) was fitted instead
 
\begin{equation} \label{eq:weibull}
  y = \frac{k}{\lambda}\bigg(\frac{M_{range}}{\lambda}\bigg)^{(k-1)}e^{-(M_{range}/\lambda)^k}
\end{equation} 

where $k$ is the shape parameter, $\lambda$ is the scale parameter and $M_{range}$ is the range of absolute magnitude values over which the fit is made. As the Weibull distribution is only valid over a positive interval, we add a constant to transform the negative absolute magnitudes to positive values

\begin{equation} \label{eq:mod}
  M_{mod} = M_{range}+M_{max}+0.1
\end{equation} 

where $M_{mod}$ is the transformed range and $M_{max}$ is the maximum value in the fit range. 

Both distribution types were fitted using a least squares curve fit in the python {\scriptsize{SCIPY}} package. The most likely absolute magnitude was the average of the Gaussian, or the mode $M_{mode}$ of the Weibull distribution, transformed back to negative values

\begin{equation} \label{eq:mode}
  M_{mode} = \lambda\bigg(\frac{k-1}{k}\bigg)^{(1/k)}-(M_{max}+0.1)
\end{equation} 

Credible intervals were again used for 68\% uncertainties on individual magnitudes. The typical variation between Monte Carlo runs (due to different data selections), was less than +/-0.05. In a small number of cases, the distribution fitting failed. In these instances, we calculated the point value of absolute magnitude, using the peaks of the distance, apparent magnitude and extinction probability curves. Due to the non Gaussian nature of the distance distributions, however, there was some offset (usually on the scale of 0.1 mag) between the peaks fitted to full distributions and these point values. 


\bibliographystyle{mnras}


\bibliography{gaia_dr2_pc} 
\makeatletter
\relax
\def\mn@urlcharsother{\let\do\@makeother \do\$\do\&\do\#\do\^\do\_\do\%\do\~}
\def\mn@doi{\begingroup\mn@urlcharsother \@ifnextchar [ {\mn@doi@}
  {\mn@doi@[]}}
\def\mn@doi@[#1]#2{\def\@tempa{#1}\ifx\@tempa\@empty \href
  {http://dx.doi.org/#2} {doi:#2}\else \href {http://dx.doi.org/#2} {#1}\fi
  \endgroup}
\def\mn@eprint#1#2{\mn@eprint@#1:#2::\@nil}
\def\mn@eprint@arXiv#1{\href {http://arxiv.org/abs/#1} {{\tt arXiv:#1}}}
\def\mn@eprint@dblp#1{\href {http://dblp.uni-trier.de/rec/bibtex/#1.xml}
  {dblp:#1}}
\def\mn@eprint@#1:#2:#3:#4\@nil{\def\@tempa {#1}\def\@tempb {#2}\def\@tempc
  {#3}\ifx \@tempc \@empty \let \@tempc \@tempb \let \@tempb \@tempa \fi \ifx
  \@tempb \@empty \def\@tempb {arXiv}\fi \@ifundefined
  {mn@eprint@\@tempb}{\@tempb:\@tempc}{\expandafter \expandafter \csname
  mn@eprint@\@tempb\endcsname \expandafter{\@tempc}}}


\makeatother



\bsp	
\label{lastpage}